\documentclass[12pt,letterpaper]{JHEP3}
\pdfoutput=1
\usepackage{epsfig,amsfonts,amssymb}
\usepackage{amsmath}
\usepackage[numbers, square, comma, sort&compress]{natbib}

\newcommand{\ben}{\begin{eqnarray}\displaystyle}
\newcommand{\een}{\end{eqnarray}}
\newcommand{\be}{\begin{equation}}
\newcommand{\ee}{\end{equation}}
\newcommand{\lb}{\left (}
\newcommand{\rb}{\right )}
\newcommand{\ltb}{\left [}
\newcommand{\rtb}{\right ]}
\newcommand{\ra}{\rightarrow}

\newcommand{\nn}{\nonumber}

\newcommand{\tran}{transformation}

\newcommand{\rn}{Reissner-Nordstr\" om}
\newcommand{\bc}{\begin{center}}
\newcommand{\ec}{\end{center}}

\newcommand{\cp}{{\ensuremath{\mathbb{CP}^2}}}
\newcommand{\cpt}{{\ensuremath{\mathbb{CP}^3}}}

\newcommand{\tst}{{\em TsT}}

\newcommand{\sch}{Schr\" odinger}
\newcommand{\nh}{near-horizon}

\def\lfig#1#2#3#4{
 \begin{figure}[h]
 \refstepcounter{figure}
 \label{#4}
 \addtocounter{figure}{-1}
 \epsfxsize=#3
 \centerline{\epsfbox{#2}}
 {\bf \caption{{\rm #1}}}
 \end{figure}
}

\def\calp         {{\cal P}}

\def\cg2{\cos (\pi V)}
\def\sg2{\sin (\pi V)}
\def\cb2{\cos (\delta/2)}
\def\sb2{\sin (\delta/2)}
\def\sa{r_0^2 {\rm sinh}^2\alpha }
\def\sg{r_0^2 {\rm sinh}^2\gamma }
\def\ss{r_0^2 {\rm sinh}^2\sigma }

\def\cg{r_0^2 {\rm cosh}^2\gamma }

\def\[{\left [}
\def\]{\right ]}
\def\({\left (}
\def\){\right )}

\title{Geometry and Phase Structure of Non-Relativistic Branes }

\author{Nabamita Banerjee$^{(a)}$\footnote{N.Banerjee@uu.nl}, Suvankar
  Dutta$^{(b)}$\footnote{pysd@swan.ac.uk} and Dileep
  P. Jatkar$^{(c)}$\footnote{dileep@hri.res.in}\\
$^a$ ITF, Utrecht University, Utrecht, The Netherlands\\
$^b$ Department of Physics, Swansea University, Swansea, UK\\
$^c$ Harish-Chandra Research Institute, Chhatnag Road, Allahabad, India}

\abstract{We use the solution generating technique, the {\em TsT}
  transformation, to obtain new solutions in type II string theory as
  well as in M-theory.  We explicitly work out examples starting with
  rotating D3 and M2 branes as well as D1-D5-p system.  Among a
  variety of solutions, we find many of them have asymptotic Schr\"
  odinger symmetry.  We also device a new method of deriving free
  energy of black brane systems, which is more efficient than the
  Euclidean action procedure.  We test our method on known examples
  before applying it to new asymptotically Schr\" odinger
  backgrounds.  We study phase structure of these backgrounds by
  analysing the free energy thus derived.}


\preprint{}

\begin{document}{\vskip 1cm}

\tableofcontents

%%%%%%%%%%%%%%%%%%%%%%%%%%%%%%%%%%%
%%%%%%%%%%%%%%%%%%%%%%%%%%%%%%%%%%%
%%%%%%%%%%%% INTRODUCTION %%%%%%%%%
%%%%%%%%%%%%%%%%%%%%%%%%%%%%%%%%%%%

\section{Introduction and Summary}

AdS/CFT correspondence relates conformal field theories in $d$
spacetime dimensions to superstring/M-theory compactified down to
anti-de Sitter spacetime in $d+1$ dimensions\cite{maldacena, magoo}.
This correspondence has been extended to non-conformal field theories
by relating it to asymptotically AdS spacetimes in the bulk.  It has
also been generalized to non-conformal branes\cite{skenderis}. 

Another direction in which this correspondence is extended is in the
exploration of holographic duals of strongly coupled non-relativistic
conformal field theories\cite{son, bm, Goldberger, abm, rangamani,
  bmw,mmt,Bobev}.  These non-relativistic conformal field theories
have Schr\" odinger group symmetry\cite{yamada, kim-yamada, abdr, 
  mukund-son, singh, Singh2, Brattan, Panigrahi, ayan}. This symmetry
consists of the usual Galilean invariance, the scaling symmetry as
well as the particle number symmetry\footnote{Relation of Schr\"
  odinger symmetry and the Newton-Cartan theory dates back to work of
  Duval et al. \cite{Duval:1984cj}.  For other related works, see,
  \cite{Duval:1990hj, Duval:2008jg, Duval:2009vt}.}.  There has been
some activity 
along this direction where solution generating techniques have been
used to obtain bulk geometries with asymptotic Schr\" odinger
symmetry\cite{mmt, theisen, imeroni-sinha, pal, pal2}.  A large class
of asymptotically \sch\ geometries have been obtained by applying
these techniques to various brane
configurations\cite{theisen,imeroni-sinha}. 

Interplay between the \tst\ transformation and the near horizon
extremal limit has been studied earlier in the case of spinning D3
brane problem\cite{imeroni-sinha}.  It was shown that the procedure of
near horizon extremal limit does not commute with the \tst\
transformation in case of both null \tst\ as well as space-like \tst\
transformation of spinning D3 brane solutions.  In this paper, we will
study \tst\ transformations of D-brane systems and their near horizon
extremal limit.  In section 2, we will write down general properties
of \tst\ transformations and list relevant formulae which will be
useful in subsequent sections.  In section 3, we will re-examine the
spinning D3 brane case both for null and space-like \tst\ \footnote{We
  will define null and space-like \tst \ transformations in the next
  section.}.  We find that, our results are in agreement with earlier
results in case of null \tst\ transformations but contrary to earlier
claims we find that space-like \tst\ transformation does commute with
the near horizon extremal limit.  Later in the same section we will
look at transformation of the D1-D5-p system with respect to null and
space-like \tst .  We find that near horizon extremal limit of this
system for null \tst\ is independent of the parameter $\gamma$, which
appears in the \tst\ transformation.  As a result, near horizon
extremal limit of D1-D5-p system before and after null \tst\
transformation gives rise to identical background. 

We then proceed to apply this method to the rotating M2 brane
solution\footnote{Similar Schr\" odinger backgrounds from M-theory
  were considered earlier in \cite{Yavartanoo, Ooguri-Park}.} in
section 4.  Since T-duality is not a symmetry of M-theory, we can
implement the \tst\ transformation by first reducing the M2-brane
solution to type IIA theory and then carrying out the \tst\
transformation on it.  We then oxidize the resulting solution back to
M-theory.  There are a couple of ways of reducing the M2-brane to type
IIA string theory solution.  One of them gives rise to fundamental
string solution whereas the other gives the D2-branes solution.
Action of \tst\ transformation on these two solutions gives different
backgrounds.  In case of fundamental string solution, null \tst\
transformation gives rise to a background which is not asymptotically
\sch\ type geometry.  The scalar $\mathcal{M}$, which parametrizes
scale factor in front of light cone part of the metric as well as
behavior of the dilaton, does not go to constant asymptotically.  In
order to get better behaved solution we work in a suitable type IIB
frame, which can be obtained by doing a T-duality on type IIA
background and followed by an S-duality.  In this type IIB frame we
carry out the \tst\ transformation.  We then revert back to the
M-theory metric in two different ways.  We can either follow same
steps in reverse order, {\em i.e.}, doing S-duality, followed by
T-duality to type IIA theory and then lift to M-theory or we can
directly T-dualize the background in to type IIA back ground and lift
it the M-theory.  These two procedures give rise to two different
M-theory solutions.  In the former case we end up with a geometry
which has a decoupled metric piece corresponding to
$\mathcal{CP}^3\times S^1$ and the full metric asymptotically
approaches $AdS_3\times\mathcal{CP}^3\times S^1$.  In the latter case,
we have decoupled $\mathcal{CP}^3$ metric and the circle is fibered
over the three dimensional non-compact space.  This space
asymptotically becomes a circle fibration over $AdS_3$ with a
decoupled $\mathcal{CP}^3$ metric.

In section 5, we discuss the phase structure of the \tst \ transformed
black holes.  The phase structure of the black hole spacetime can be
obtained from its free energy.  Therefore one needs to compute the
free energy from the Euclidean path integral of black hole spacetime,
\be
F = -T \ln Z_S\,,
\ee
where, $Z_S$ is the partition function. For pure gravity the partition
function $Z_S$ is given by    
\ben \label{Zs}
Z_{S}(X) = \int \left [ {\cal D}g \right ] e^{- I_{S}(X)}\,,
\een
where, $I_S(X)$ is the Euclidean effective action on  $X$.  In
large $N$ limit (or $G_N \ra 0$, classical supergravity)
it is given by\footnote{We are considering only
gravity for simplicity.},
\ben
I_S &\sim& \frac{1}{G_N}\int d^{d+1}x \sqrt{g
} \ R \\ \nonumber
&=& N^2 K(X).
\een
In general, there may be several possible X's given a fixed boundary
geometry $M$.  In that case there is no particular way to pick one
specific spacetime.  Therefore, one has to consider all possible X's
and replace (\ref{Zs}) by sum over all such integral.  In semi-classical
limit, {\it i.e.} $N\ra \infty$, one can write the partition function
in the following way,
\be \label{Zssemi}
Z_{S}(X) =  \sum_{i}e^{-N^2
    K^{cl}(X^0_i)}\,,
\ee
where, $X^0_i$'s are classical solutions of the Einstein equation with
some fixed boundary geometry.

One can notice that there is a natural mechanism for a singularity or
phase transition that would arise only in the large $N$ limit.  At
large $N$, the sum will be dominated by those $X^0_i$ for which
$F(X_i^0)$ is smallest.  Therefore at a point at which $F(X_i^0) =
F(X_j^0)$ for some $i \neq j$, one may jump from one branch to other
branch.  This signifies a phase transition in the large $N$ theory.

To study the phase transition we need to compute the on-shell action.
In this paper we develop a new method to compute the Euclidean on-shell
action.  In the usual way of computing on-shell action one encounters large
volume divergence and one needs to regularize the action by either subtracting
the contribution from background spacetime or adding counterterms to
the original action.  In practice, these methods are cumbersome especially
when one has several matter fields.  We use Wald's formula to compute 
entropy of the black hole which depends only on the near-horizon geometry.
We derive temperature of the Euclidean black hole by computing the
periodicity of Euclidean time circle.  We then use these
two variables to compute the on-shell action of the black hole
using the following relations, 
\ben
S-\beta \lb {\partial I \over \partial \beta} \rb_{\mu} + I &=&0 \qquad
{\rm for\ fixed\ chemical\ potential} \ \mu \,,\nn\\
S-\beta \lb {\partial I \over \partial \beta} \rb_{q} + I &=& 0 \qquad
{\rm for\ fixed\ charge} \ q\, .
\een
When we integrate these first order equations we encounter an integration
constant.  We fix this constant by demanding the free energy of the background
spacetime to be zero.  We describe this procedure in detail in section
(\ref{thermo}), and work out an illustrative example before applying
it to the non-relativistic case.  The Euclidean on-shell action
derives in this manner for black hole spacetimes is then used to study
phase structure.  We will conclude with discussion of our results.

\section{The \tst\ Transformation}

In this section we will discuss general properties of \tst\
transformations, which is a solution generating technique in string
theory.  The basic idea of solution generating technique is to exploit
the fact that the low energy supergravity theory has more symmetry
than the full string theory.  A solution to supergravity equations of
motion can be shown to be a solution to string theory.  A symmetry
transformation of this solution with respect to a supergravity
symmetry does not give rise to a new solution in supergravity.
However, if the transformation employed is not a symmetry of full
string theory then the transformed solution can be interpreted as a
new solution of string theory.  In case of the \tst\ transformations,
T-dualities are performed along the isometry direction, whereas {\em
  s}-transformation mixes the dualized coordinate with another
isometry direction.  This transformation generically changes
asymptotics of the resulting metric but is guaranteed to be a solution
to the supergravity equations of motion.  We will spell out the
procedure involved in \tst\ transformation shortly but before that, we
will briefly mention two metrics, which give rise to non-relativistic
isometries in the boundary conformal field theories and appear in the
\tst\ transformed solutions.

The holographic dual of a $d$ spatial dimensional Galilean CFT is
given by \cite{son, bm, abm},
\be\label{sch}
ds^2 = r^2 \lb -2 du dv -r^{2(z-1)} du^2 + \sum_{i=1}^d (dx^i)^2 \rb
+\frac{dr^2}{r^2}.
\ee
For $z=1$ the spacetime becomes pure AdS$_{d+3}$ whose
dual theory corresponds to a conformal field theory in $(d+1,1)$
spacetime dimensions.

For $z>1$ the the bulk geometry corresponds to a dual to a
non-relativistic field theory with the Galilean transformations along
with scaling as a global symmetry.  While the coordinate $u$ is
interpreted as a boundary time coordinate, the null direction $v$ is
taken to be compact and the quantized momentum along $v$ is identified
with the particle number of Galilean CFT.  We call this spacetimes
Sch$^z_{d+3}$, where, $z$ is the dynamical exponent of the \sch\ spacetime.

Let us also note that the bulk geometry corresponding to
Lifshitz spacetime is given by,
\be
ds^2 = r^2 \lb- r^{2(z-1)} dt^2 + \sum_{i=1}^d (dx^i)^2\rb + \frac{dr^2}{r^2},
\ee
where $t$ is the boundary time coordinate of the boundary theory as well.

We will now discuss the generic form of the \tst\ transformed
geometry.  We will use the unhatted variables to denote the metric,
dilaton and the second rank anti-symmetric tensor before \tst\
transformation and hatted variables will denote them after \tst .  In
case of Ramond-Ramond(RR) field strengths our convention is to use
$F_q$ to denote $q$-form before \tst\ and $\mathcal{F}_q$ to denote
them after \tst .  In the next section, we will apply this technique
to specific examples in string theory.  The \tst\ transformations can
be applied to those metrics which have at least two isometry
directions.  In absence of NS-NS two form field B, it
 is convenient to cast the metric in the following form
\begin{equation}\label{easyform}
	ds^2 = ( A_1 d\chi + K_1)^2 + (A_2 d \psi + A_3 d\chi + K_2)^2 + ds^2_8\,,
\end{equation}
where $ds^2_8$ and the one-forms $K_1$, $K_2$ do not depend on the
isometry directions $\chi$ and $\psi$.  The \tst\ transformations then
correspond to performing T-duality along $\psi$ direction, followed by
a shift along $\chi$, {\it i.e.}, $\chi \to \chi - \gamma \psi$, and
then performing T-duality back along the $\psi$ direction.  Notice
that under the Buscher duality transformations\cite{Buscher}, pure
metric background can give rise to non-trivial dilaton and
anti-symmetric tensor field background.  For example, if the original
metric is of the form (\ref{easyform}) then the end result of the
\tst\ transformation is
\begin{eqnarray}\label{NSNSTsT}
	d\hat s^2 &=& {\cal M} ( A_1 d\chi + K_1)^2 + {\cal M} (A_2 d
        \psi + A_3 d\chi + K_2)^2 + ds^2_8\,,\nn \\
	e^{2 \hat\Phi} &=& {\cal M} e^{2 \Phi}\,,\quad
        {\cal M} = (1 + \gamma^2 A_1^2 A_2^2)^{-1}\,, \nn\\
	\hat B &=& - \gamma {\cal M}\, A_1 A_2\, ( A_1 d\chi + K_1)
        \wedge (A_2 d \psi + A_3 d\chi + K_2)\,.
\end{eqnarray}

In this simplified situation, where $B=0$ and only $p$-form RR field
strength is non-vanishing, we can easily find the effect of \tst\ on
the RR field.  Suppose this $p$-form field strength has
components along both $\psi$ and $\chi$ direction, then after \tst\
transformation we get $(p-2)$-form field strength\
\begin{equation}
  \label{eq:2}
  \mathcal{F}_{p-2} = \gamma \iota_{\chi}\iota_{\psi} F_p\, ,
\end{equation}
in addition to the original $p$-form field strength.  As a result, the
effective $p$-form field strength is then given by
\begin{equation}
  \label{eq:3}
   F_p = \mathcal{F}_p + \mathcal{F}_{p-2}\wedge \hat B\, .
\end{equation}

In general, original background can have non-trivial dilaton as well
as antisymmetric field strength. To find the \tst\ transformed form of
the metric and the NS-NS field strength it is useful to define \cite{imeroni},
\be
E_{\mu\nu} = g_{\mu\nu}+B_{\mu\nu}.
\ee
Under \tst \ transformation, $E_{\mu\nu}$ transforms as,
\begin{equation}
  \label{eq:4}
  \hat E_{\mu\nu} = {\cal M} \ltb E_{\mu\nu} +\gamma \ltb \det \left(
    \begin{array}[h]{cc}
      E_{\psi\chi} & E_{\psi\nu}\\
      E_{\mu\chi} & E_{\mu\nu}
    \end{array}
  \right) -
  \det \left(
    \begin{array}[h]{cc}
      E_{\chi\psi} & E_{\chi\nu}\\
      E_{\mu\psi} & E_{\mu\nu}
    \end{array}
  \right)\rtb
 + \gamma^2 \det \left(
    \begin{array}[h]{ccc}
      E_{\psi\psi} & E_{\psi\chi} &  E_{\psi\nu}\\
      E_{\chi\psi} & E_{\chi\chi} & E_{\chi\nu}\\
      E_{\mu\psi} & E_{\mu\chi} & E_{\mu\nu}
    \end{array}
  \right)\rtb
\end{equation}
where,
\be
{\cal M} = \ltb 1 + \gamma (E_{\psi\chi}-E_{\chi\psi}) + \gamma^2 \det
 \left(
    \begin{array}[h]{cc}
      E_{\psi\psi} & E_{\psi\chi} \\
      E_{\chi\psi} & E_{\chi\chi}
    \end{array}
  \right)\rtb^{-1}.
\ee
We can now read out the transformed metric and NS-NS
antisymmetric field  from the equation (\ref{eq:4}) by isolating
symmetric and antisymmetric parts of $\hat E$, namely,
\ben
\hat g_{\mu\nu} = {\rm Sym}[\hat E_{\mu\nu}]\, ,\nn \\
\hat B_{\mu\nu} = {\rm AntiSym}[\hat E_{\mu\nu}]\, .
\een
The dilaton transforms as,
\be
 e^{2 \hat\Phi} = {\cal M}\ e^{2\Phi}.
\ee
The \tst\ transformation also acts on the RR field strengths.  For
general backgrounds, RR fields and the field strengths after \tst\ are related to
those before \tst\
\begin{eqnarray}
  \label{eq:1}
\sum_q \hat C_q \wedge \exp(\hat B) &=& \sum_q C_q \wedge \exp(B)
  + \gamma \iota_\chi \iota_\psi \sum_q C_q \wedge \exp(B)\, , \nn \\
  \sum_q \mathcal{F}_q \wedge \exp(\hat B) &=& \sum_q F_q \wedge \exp(B)
  + \gamma \iota_\chi \iota_\psi \sum_q F_q \wedge \exp(B)\, ,
\end{eqnarray}
where $B$ is the Neveu-Schwarz sector antisymmetric tensor before
\tst\ transformation and $\hat B$ is that after \tst . In the next two
sections, we will
apply these general \tst\ transformation rules to different string
theory and M-theory geometries to generate deformed solutions.

\section{\tst \ Transformation of String Geometry }

In this section we discuss the effect of \tst \ \tran \ of $Dp$ brane
geometry and study how their asymptotic and near horizon geometries
change under this \tran . In particular we consider two different $D$
brane configurations, rotating $D3$ brane geometry and $D1$-$D5$-$p$
system.  Both arise as solutions of type IIB string theory.  We perform
null as well as space-like \tst \ \tran s in both the cases and
discuss the behavior of geometries asymptotically as well as in the near
horizon limit. We also study the commutativity of extremal near horizon
limit and \tst \ \tran \ in both the cases.

\subsection{Rotating $D3$ Brane}

We start with a review on the non-relativistic extension of the
rotating non-extremal $D3$-brane solution.  Let us consider $D3$
branes rotating along three isometry directions of transverse $S^5$
space with equal angular momenta in all three planes.  From five
dimensional point of view (compactifying over $S^5$) this system can
be viewed as a charged $AdS_5$ black brane with $U(1)^3$ symmetry.
The boundary theory corresponds to a conformal field theory with three
equal global $U(1)$ charges.  This system has been studied in details in
\cite{imeroni-sinha}.  In most of the cases our results are in
agreement with theirs, however, there are some differences in case of
space-like \tst\ case.

We will be working in ten dimensional set-up and we will analyze
effect of both Null \tst \ and space-like \tst \ transformations.
Since \tst \ transformation is a symmetry of supergravity, the
non-relativistic geometries resulting out of these transformations are
solutions of type IIB theory.  In the end we will take extremal near
horizon limit of both \tst \ transformed solutions.  As mentioned in
the introduction, our results agree with earlier results in the
literature about non-commutativity of null \tst\ transformation and
the near horizon extremal limit.  In case of space-like \tst ,
however, we find that it commutes with the near horizon extremal
limit.  The non-extremal metric\footnote{We follow the notation of
  \cite{imeroni-sinha}.}  in the decoupling limit \cite{maldacena} of
the rotating $D3$ brane geometry is given by,
\begin{eqnarray}
 \label{BH}
ds^2 &=& \frac{r^2}{l^2} \left( - f dt^2 + dx^2 + dy^2 + dz^2 \right)
		+ \frac{l^2}{r^2} \frac{dr^2}{f}\\
		&& + l^2 \left( d\alpha^2 + \sin^2\alpha d\beta^2
		+ \mu_1^2 (d \xi_1 + A)^2 + \mu_2^2 (d \xi_2 + A)^2 +
                \mu_3^2 (d \xi_3 + A)^2 \right)\,,\nn \\
	F_5 &=& (1 + *) \left[ \left( -\frac{4 r^3}{l^4} dt \wedge dr
		+\frac{Q}{l^2}\, d \left( \sum_{i=1}^3 \mu_i^2 d \xi_i \right)
		 \right) \wedge dx \wedge dy \wedge dz \right] \,,
\end{eqnarray}
where the angular functions $\mu_i$ are parameterized as,
\begin{equation}
	\mu_1 = \cos \alpha\,,\qquad
	\mu_2 = \sin \alpha \cos \beta\,,\qquad
	\mu_3 = \sin \alpha \sin \beta\,,
\end{equation}
and the metric functions are as follows,
\begin{equation}\label{metfuns}
	f(r) = \left( 1 - \frac{r_0^2}{r^2} \right)
		\left( 1+ \frac{r_0^2}{r^2} - \frac{Q^2}{r_0^2 r^4} \right)\,,\qquad
	\mathcal{A} = \mathcal{A}_t\, dt = \frac{Q}{l^2} \left(
          \frac{1}{r_0^2} - \frac{1}{r^2} \right) dt \,.
\end{equation}
In what follows we will consider three angular momenta to be equal and
denote them by $Q$.

As mentioned in the previous section we will need two isometry
directions to carry out the \tst\ transformation.  We will use one of
the D3-brane world volume direction as an isometry direction.
Second isometry direction that we will use will be from $S^5$
direction.  To extract that it is suitable to write the $S^5$ metric
as a U(1) fibration over $\cp$,
\ben
ds_{S^5}^2=  \lb d\psi + \calp + \mathcal{A}\rb^2 + ds_{\rm{CP^2}}^2\,,
\een
where the one form $\cal{A}$ is defined in (\ref{metfuns}) and,
\be
\calp = \frac{1}{3} \left(\text{d$\chi $}_1+\text{d$\chi
   $}_2\right)-\sin ^2\alpha \left(\text{d$\chi
   $}_2 \sin ^2\beta+\text{d$\chi $}_1 \cos
   ^2\beta\right),
\ee
\be
\psi=\frac13(\phi_1+\phi_2+\phi_3),\ \
\chi_1= \phi_1-\phi_2, \ \
\chi_2= \phi_1-\phi_3\, .
\ee
Note that curvature of the one form $\cal{P}$ is proportional to the
K\" ahler form $\omega_{\cp}$ on $\cp$, namely, $-\frac{l^2}{2} d{\cal
  P}=\omega_{\cp}$ and the metric of
$\cp$ can be expressed as,
\ben
ds_{\rm{CP^2}}^2 &=& d\alpha^2 + \sin^2\alpha d\beta^2 \nn\\
&& + \sin^2 \alpha\cos^2\alpha(\cos^2\beta d\chi_1 + \sin^2\beta
d\chi_2)^2\nn\\
&& +  \sin^2\alpha \sin^2 \beta \cos^2 \beta (d\chi_1-d\chi_2)^2\, .
\een
Thus, the rotating D3 brane solution that we are looking at has all three
angular momenta equal.  In the next sub-sections we will study the
non-relativistic extension of this geometry.

\subsubsection{Null \tst \ Transformation}

We will start with the null \tst\ transformation.  Before performing the
 transformation, we first need to define light-cone co-ordinates,
\be
x^{\pm}=\frac{1}{2}(t\pm y).
\ee

The vielbeins $e^i$, which are derived from the original metric,
are given by,
\ben
e^0 &=& \frac{2 r \sqrt{f}}{\sqrt{1-f}} dx^+, \quad e^1 = {r \ltb
  (1-f)dx^- -(1+f)dx^+ \rtb  \over l \sqrt{1-f}}\,,\quad
e^2 = \frac r l dx\,,\quad e^3 = \frac r l dz, \nn\\
 e^4 &=&  {dr \over r
  \sqrt{1-f}}\,,\quad
 e^5 = l d\alpha\,, \quad e^6=l \sin\alpha d\beta\,,\quad
e^7 = \frac{l}{2} \sin 2\alpha (\cos^2\beta d\chi_1 + \sin^2\beta
d\chi_2)\,,\nn\\
e^8 &=& l \sin\alpha \sin\beta \cos\beta (d\xi_1 - d\xi_2),\quad
e^9=l (d\psi+ \mathcal{P} +\mathcal{A}).
\een
To connect up to the notation of sec.2, we will use $x^-$ as $\chi$
direction in the \tst\ transformation described there.  In the
light-cone coordinates, different functions appearing in the metric
(\ref{easyform}) take the following form:
\ben
A_1 &=& \frac{r}{l} \sqrt{1-f}, \qquad A_2= l\,, \quad A_3=\mathcal{A}_t\,,
\qquad K_1=- \frac{r}{l} \frac{(1+f)}{\sqrt{1-f}}dx^+\,, \nn\\
K_2 &=& \frac{1}{3}(d\chi_1+d\chi_2) - \sin ^2 \alpha
\left(d\chi_1 \cos ^2\beta + d\chi_2 \sin ^2\beta \right)+
\mathcal{A}_t dx^+\,.
\een
Notice that the NS-NS two form $B$ is zero for the D3 brane solution
(\ref{BH}).  Following the general \tst\ transformation procedure
outlined in sec.2, we write down the new solution obtained from
eq.(\ref{BH}).  The \tst\ procedure is implemented after expressing
(\ref{BH})  in terms of the lightcone coordinates, $x^\pm$.  The new
background, which also solves type IIB equations of motion, is
\ben\label{NTST}
d\hat s^2 &=& {\cal M} ( A_1 dx^- + K_1)^2 + {\cal M} (A_2 d \psi +
A_3 dx^- + K_2)^2 + ds^2_8\,,\nn\\
	e^{2 \hat\Phi} &=& {\cal M} = \left( 1 + \gamma^2 r^2 (1-f)
        \right)^{-1}\,,\quad 	\hat C_2 = - \gamma l^2 \mathcal{A}_t \,
        \omega_{\cp}\,, \nn\\
	\hat B &=& -\gamma r {\cal M}\sqrt{1-f} ( A_1 dx^- + K_1)
        \wedge (A_2 d \psi + A_3 dx^- + K_2)\,, \\
	\mathcal{F}_5 &=& F_5 + \hat H_3 \wedge \hat C_2 = (1 + *)\, G_5\nn\\
		& =& (1 + *) \Big[ -\frac{4}{l}\, e^0 \wedge e^1
                \wedge e^2 \wedge e^3 \wedge e^4
		+\frac{2 Q}{l r^3 \sqrt{1-f}}\, (\sqrt{f} e^0 + e^1)
		\wedge e^2 \wedge e^3 \wedge \omega_{\cp} \Big]\,, \nn
\een
The Poincare dual is defined using  $*$ is defined with respect to this \tst\
transformed metric.  We can also write the deformed geometry
(\ref{NTST}) in terms of original $(t,y)$ coordinates as,
\ben\label{NTsT}
d\hat s^2 &=&\frac{r^2}{l^2} {\cal M}
		\left( - f dt^2 + dx^2 - \gamma^2 r^2 f (dt + dy)^2 \right)
		+ \frac{r^2}{l^2} \left(dx^2 + dz^2 \right) +
                \frac{l^2}{r^2} \frac{dr^2}{f}\nn \\
		&&\qquad\, +\ l^2 \left( {\cal M}\, (d\psi +
                  \mathcal{P} + \mathcal{A})^2 + ds^2_{\cp} \right) \,, \\
e^{2 \hat\Phi} &=& {\cal M}\,, \quad
	\hat B =  \gamma r^2 {\cal M}\, (f dt + dy) \wedge (d\psi +
        \mathcal{P} + \mathcal{A})\,,\quad
	\hat C_2 = - \gamma l^2 \mathcal{A}_t \, \omega_{\cp} \,,\nn \\
	\mathcal{F}_5 &=& (1 + *)\, G_5 = (1 + *) \left[ \left(
            -\frac{4 r^3}{l^4} dt \wedge dr
		- \frac{2Q}{l^4}\, \omega_{\cp} \right) \wedge dx
              \wedge dy \wedge dz \right]. \nn
\een
Let us now look at the asymptotic geometry ($r\to \infty$) of the
above solution. For this we will revert back to the light cone variables,
\ben
d\hat s^2 &\ra& \frac {r^2}{l^2} \lb -4\gamma^2 r^2 (dx^+)^2- 4dx^+dx^- +
dx^2 +dy^2\rb\nn \\
&&  +\frac{l^2}{r^2}\frac {dr^2}{r^2} + l^2 \lb (d\psi +
{\cal P} + \mathcal{A}_{\infty})^2 + ds_{\cp}^2 \rb\,,\nn\\
\exp(2\hat \Phi) &\ra& 1, \quad \hat B \ra \gamma\, r^2\, dx^+ \wedge (d\psi
+{\cal P}+\mathcal{A}_{\infty})\,,\nn\\
\hat C_2 &\to & -\gamma\, l^2\, \mathcal{A}_{t\infty}\, \omega_{\cp}.
\een
Defining rescaled light-cone coordinates as,
\be\label{d3}
u= 2 \gamma l x^+ ,\qquad v=\frac{1}{\gamma l}x^-\,,
\ee
it is easy to see that the above metric takes the form (\ref{sch}) at
the boundary $r \to \infty$ with dynamical exponent $z=2$.

\noindent
{\underline{\bf Extremal Near-Horizon Limit}}
%\bigskip

We are interested in studying the near horizon extremal limit of
this deformed geometry.  As is well known, that the undeformed
geometry (\ref{BH}) has a $AdS_2$ factor in its extremal near-horizon
limit.  The extremal limit is arrived at by taking, $Q=\sqrt{2} r_0^3$.
Taking near-horizon limit of the extremal geometry is little subtle
and usually one needs to do an appropriate scaling of coordinates to
get to the near horizon geometry.  We will consider the following scaling
limit,
\be\label{nhscalling}
\rho= \frac{r-r_0}{\beta}, \qquad \tau=\beta t,\quad {\rm with} \quad
\beta \ra 0.
\ee
This gives the near-horizon extremal geometry as,
\ben\label{TsTnh}	
d\hat s^2 &=&\tfrac{r_0^2}{l^2}
		\left( - \tfrac{12 \rho^2}{r_0^2} dt^2 + {\cal M}_0\, dy^2 \right)
		+ \tfrac{r_0^2}{l^2} \left(dx^2 + dz^2 \right) +
                \tfrac{l^2}{12 \rho^2} d\rho^2  \nn \\
		&&\qquad\qquad + l^2 \left({\cal M}_0\, \left(d\psi +
                    P + \tfrac{2\sqrt{2}\rho}{l^2} dt \right)^2
		+ ds^2_{\cp} \right)\,,\\
	e^{2 \hat \Phi} &=& {\cal M}_0\,,\quad
	\hat B = \gamma r_0^2\, {\cal M}_0\, dy \wedge
		\left(d\psi + P + \tfrac{2\sqrt{2}\rho}{l^2} dt
                \right)\,,\quad \hat C_2 = 0\,, \nn \\
	{\cal F}_5 &=& (1 + *) \left[ \left( -\tfrac{4 r_0^3}{l^4} dt \wedge d\rho
		- \tfrac{2\sqrt{2}r_0^3}{l^4}\, \omega_{\cp} \right)
              \wedge dx \wedge dy \wedge dz \right]\,, \nn
\een
where, ${\cal M}_0 = (1+\gamma^2 r_0^2)^{-1}$.

Thus we see that, while the asymptotic geometry has \sch\ isometry, we
do get the $AdS_2$ factor in the near horizon extremal limit.  The
same geometry (\ref{TsTnh}) can also be obtained, if we start from the
extremal near horizon solution (which has an $AdS_2$ factor in it),
perform null \tst \ and then take the above scaling limit
(\ref{nhscalling}).  It is worth mentioning that in this case before
taking the scaling limit, the near horizon geometry does not have any
$AdS_2$ isometry, as pointed out by \cite{imeroni-sinha}.

\subsubsection{Space-Like \tst \ Transformation}

For space-like \tst \ \tran \ we do not use the light-cone
coordinates, we instead choose $y$ as $\chi$ direction and perform the
twist in $y$ direction. The deformed geometry now looks like:
\ben\label{STsT}
d\hat s^2 &=&\frac{r^2}{l^2}
		\left( - f dt^2 \right)+{\cal M} \frac{r^2}{l^2} dy^2
		+ \frac{r^2}{l^2} \left(dx^2 + dz^2 \right) +
                \frac{l^2}{r^2} \frac{dr^2}{f}\nn \\
		&&\qquad + l^2 \left( {\cal M}\, (d\psi +
                  \mathcal{P} + \mathcal{A})^2 + ds^2_{\cp} \right)\,, \nn \\
e^{2 \hat\Phi} &=& {\cal M} \,,\quad
	\hat B = - \gamma r^2 {\cal M}\,  dy \wedge (d\psi +
        \mathcal{P} + \mathcal{A})\,,\quad
	\hat C_2 = - \gamma l^2 \mathcal{A}_t \, \omega_{\cp}\,,\nn \\
	\mathcal{F}_5 &=& (1 + *)\, G_5 = (1 + *) \left[ \left(
            -\frac{4 r^3}{l^4} dt \wedge dr
		- \frac{2Q}{l^4}\, \omega_{\cp} \right) \wedge dx
              \wedge dy \wedge dz \right]\,,
\een
where, ${\cal M}=\left( 1 + \gamma^2 r^2  \right)^{-1} $.

The \nh \ extremal limit gives similar result as above, only with an
opposite sign for the two-form field $\hat B$. It also solves the
equations of motion as expected. For both cases, we see that the \nh \
extremal geometry does depend on the shift parameter $\gamma$.  We
will see that the situation is different in the $D1$-$D5$-$p$
system that we will study in the next section. Starting with extremal
\nh \ geometry and a space-like \tst \ and then scaling gives back the
above geometry.

Thus we see that the extremal \nh \ limit of \tst \ transformed
geometry and the \tst \ transform of near horizon extremal geometry are $not$
same. We summarize our observations in the following (non)commutative
diagram.
\lfig{TsT transformation and \nh \ geometry.}
{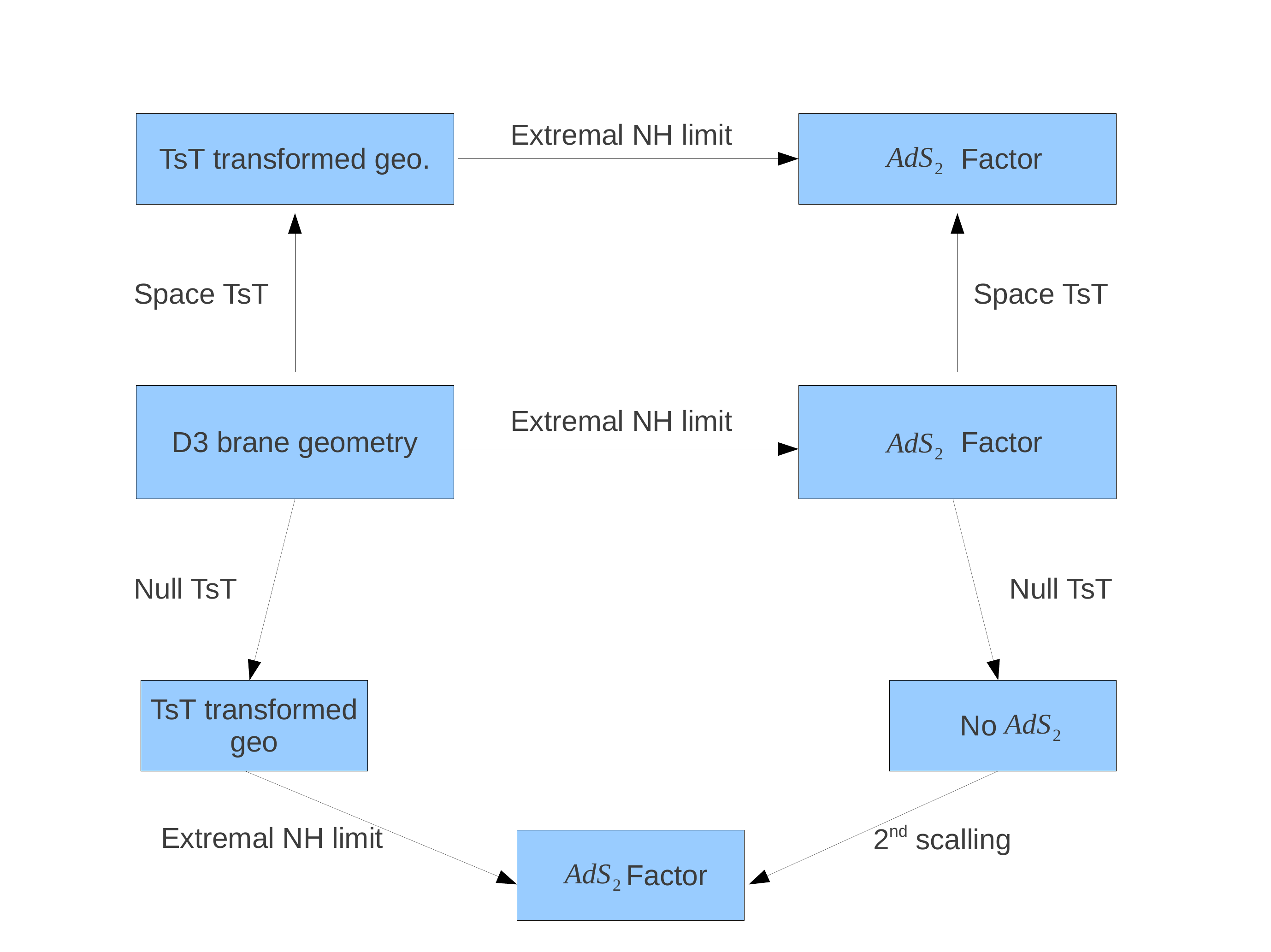}{9.5cm}{fig1}

\subsection{D1-D5-p System}

We will now look at another solution of type IIB theory.
This is given by the D1-D5 brane system with momentum $p$, where $Q_1$
number of D1 branes wrap the compact direction $x_5$, $Q_5$ D5-brane
wraps $x_5$ and a compact four manifold $M_4$ and the momentum $n$ is
along the circle $x_5$.  In the string frame, the metric, the dilaton
and the RR 3-form field strength $F_3$ corresponding to this
solution are given by\cite{Horowitz:1996ay},
\ben\label{d1d5}
ds^2_{\rm{str}}&=& {1\over \sqrt{H_1 H_2}}\bigg[ -dt^2 + dx_5^2 + {c_3
  \over r^2} (\coth\sigma dt + dx_5)^2
+H_1 dM_4^2 \bigg]\nn \\
&& +
\sqrt{H_1 H_2} \bigg[f^{-1} dr^2 + r^2 d\Omega_3^2\bigg]\,, \nn \\
e^{-2\phi } &=& \lb 1+   { \sg \over r^2 }\rb \lb 1 + {\sa\over r^2
}\rb^{-1}\,,\nn \\
F_3&=& {2\sqrt{c_1(c_1+r_0^2)}\over r^3 H_1^2}dt\wedge dx_5 \wedge dr
+ {2\sqrt{c_2(c_2+r_0^2)} \over r^3} \epsilon_3,
\een
where $\epsilon_3$ is the volume element of 3 sphere. The harmonic
functions appearing in the metric are,
\be
H_1= 1 + {c_1 \over r^2}, \ \ \ \ H_2 = 1 + {c_2 \over r^2} \ \ \ \ f=
1-{r_0^2\over r^2},
\ee
and the parameters $c_i$ are as follows,
\be
c_1= \sa, \ \ \ \ c_2= \sg, \ \ \ \
c_3 = \ss\,.
\ee
The corresponding brane charges and the momentum are expressed in
terms of $c_i$, the volume of $M_4$, $V$, the radius of $x_5$ circle,
$R$ and the string coupling $g$:
\ben
Q_1 &=& {V \over 4\pi^2 g}\int * F_3 = {V \over 2 g}
2\sqrt{c_1(c_1+r_0^2)}\,, \nn\\
Q_5 &=& {1 \over 4\pi^2 g}\int F_3 = {1 \over 2 g}
2\sqrt{c_2(c_2+r_0^2)}\,, \nn\\
n&=& {R^2 V r_0^2 \over 2 g^2} \sinh2 \sigma\,.
\een

Extremal limit of this background corresponds to taking $\alpha,\,
\gamma,\, \sigma$ to $\infty$ and $r_0\to 0$ keeping $c_1,\, c_2,\,
c_3$ fixed, {\em i.e.},
\ben\label{ext}
\lim_{r_0\ra 0 \atop \alpha \ra \infty} r_0^2 \sinh^2 \alpha &\ra& c_1,\quad
\lim_{r_0\ra 0 \atop \alpha \ra \infty} r_0^2 \sinh^2 \gamma \ra c_2,\quad
\lim_{r_0\ra 0 \atop \alpha \ra \infty} r_0^2 \sinh^2 \sigma \ra c_3.
\een
We will now determine the near horizon geometry of the extremal
solution.  The near-horizon extremal geometry, as expected, has a
$AdS_2$ factor.  To obtain this, we first do the following coordinate
transformation
\be\label{nh1}
r=\sqrt{c_3 \lambda \tilde\rho}, \ \ \ x_5=y-t, \ \ \ \tau = \lambda t,
\ee
and then take $\lambda\to 0$ which corresponds to the near-horizon limit.
The near-horizon extremal metric becomes,
\ben
ds_{\rm str}^2 &=& {\sqrt{c_1 c_2}\over 4} \lb - {4 c_3 \over c_1 c_2}
\tilde\rho^2 d\tau^2 + {d\tilde\rho^2 \over \tilde\rho^2} \rb +
{c_3\over \sqrt{c_1 c_2}}(dy-\tilde\rho d\tau)^2+ \sqrt{{c_1\over
    c_2}}dM_4^2 + \sqrt{c_1 c_2} d\Omega_3^2\,.\nn\\
\een
Changing variable $\rho=2\tilde \rho  \sqrt{{c_3\over c_1c_2}}$ we get,
\ben\label{extnh}
ds_{\rm str}^2 &=& {\sqrt{c_1 c_2}\over 4} \lb - \rho^2 d\tau^2 +
{d\rho^2 \over\rho^2} \rb + {c_3\over \sqrt{c_1 c_2}}(dy-{1\over
  2}{c_1c_2\over c_3}\rho d\tau)^2+ \sqrt{{c_1\over c_2}}dM_4^2 +
\sqrt{c_1 c_2} d\Omega_3^2\,.\nn\\
\een
Thus, we get the $AdS_2$ factor with momentum along the $y$ direction.
The three form field strength in this limit is given by,
\be
F_3= {1\over 2} \sqrt{{c_2 c_3\over c_1}}d\tau \wedge dy \wedge d\rho
+ {2 c_2 \over \rho^3} \epsilon_3.
\ee

Next, we will apply both, null and space-like \tst\ transformations
on this geometry and find the deformed solutions.

\subsubsection{Null \tst \ Transformation}

 We will work with full ten dimensional geometry (\ref{d1d5}).
 We start with the null \tst\ transformation, it is then convenient to
define the null coordinates as,
\be
x^{\pm}= \frac{1}{2}(t\pm x_5).
\ee

Now, in terms of these new coordinates, we can rewrite the D1-D5 metric as,
\be
ds^2 = ( A_1 d x^- + K_1)^2 + (A_2 d \psi + A_3 d x^- + K_2)^2 + ds^2_8 ,
\ee
where,
\ben
A_1&=&\frac{\sqrt{c_3} (\coth\sigma-1)}{r(H_1 H_2)^{\frac{1}{4}}},
\,\,\,\ A_2 = r (H_1 H_2)^{\frac{1}{4}}, \,\,\,\ A_3=0,  \nn \\
K_1&=&-\frac{2 r^2- c_3 (\coth^2\sigma-1)}{r (H_1
  H_2)^{\frac{1}{4}}\sqrt{c_3}(\coth\sigma-1)}dx^+, \,\,\,\,\ K_2= r
(H_1 H_2)^{\frac{1}{4}} \frac{1}{2}\cos\theta d \phi ,
\een
and all other components of the metric are written in $ ds^2_8$.  The
$ds_8^2$ does not play any role in what follows.  The $S^3$ metric is
written as a Hopf fibration over $S^2$.  While the fibered coordinate
is written explicitly, $S^2$ metric is part of $ds_8^2$.  More concretely,
\be
r^2\sqrt{H_1 H_2} d\Omega_3^2= (A_2 d \psi + A_3 d x^- + K_2)^2+
\frac{A_2}{4}(d \theta^2+\sin^2\theta d\phi^2)\,.
\ee
As far as the NS-NS two form field $B$ is concerned the situation is
similar to the D3-brane case, namely the $B$ field is vanishing in
this background.  Thus, we can readily write down the \tst\ transformed
geometry as (\ref{NSNSTsT}),  where,
\be
{\cal M}= (1+ \gamma^2 A_1^2 A_2^2)^{-1}= (1+ \gamma^2
c_3(\coth\sigma-1)^2)^{-1}.
\ee
Again the 3-form field strength $F_3$ does not change under this
transformation.

We can rewrite the above solution in terms of original coordinates as,
\ben
d\hat s^2_{\rm{str}}&=& {{\cal M}\over \sqrt{H_1 H_2}}\bigg[ -dt^2 +
dx_5^2 + {c_3 \over r^2} ( \coth\sigma dt + dx_5)^2 + \gamma^2
(c_3(\coth^2\sigma-1)-r^2)(dt+dx_5)^2\bigg]\nn \\
&+&\sqrt{\frac{H_1}{H_2}} dM_4^2
+
\sqrt{H_1 H_2}\frac{dr^2}{f} + {\cal M}(A_2 d \psi + A_3 d x^- +
K_2)^2+\frac{B}{4}(d \theta^2+\sin^2\theta d\phi^2)\nn \\
e^{2 \hat \Phi}&=&{\cal M} \nn \\
\hat B&=&- \gamma {\cal M}\, A_2\, (c_3(\coth\sigma-1)-r^2) (
\coth\sigma dt + dx_5) \wedge (A_2 d \psi + A_3 d\chi + K_2)\,,
\een
and $F_3$ does not change its form under this transformation\footnote{In \cite{theisen} the authors found the geometry for $n=0$ case.}.

%\bigskip
\noindent
{\underline {\bf Extremal Near-Horizon Limit}}
%\bigskip

We want to study the extremal near horizon geometry of this
non-relativistic geometry. The limit is defined in (\ref{ext}) and
(\ref{nh1}). In this limit, $\tanh\sigma \to 1$ and we can easily check
that the geometry is same as the one given in (\ref{extnh}); in
particular the NS-NS two form field $B$ vanishes.  This fact has an
interesting implication on the entropy of the non-relativistic black
holes.  Since extremal near horizon geometry is identical for both
original and \tst\ transformed case implies the entropy of the
non-relativistic black hole is identical to that of the relativistic
one even after taking higher derivative corrections into account.

\subsubsection{Space-Like \tst\ Transformation}

Let us now look at the effect of space-like \tst\ transformations on
the $D1$-$D5$-$p$ system.
In this case, we will choose the two symmetry directions as, $x_5$ and
$\psi$.  We again re-write the metric as,
\be
ds^2 = ( A_1 d x_5 + K_1)^2 + (A_2 d \psi + A_3 d x^- + K_2)^2 +
d\tilde{s}^2_8 ,
\ee
where,
\ben\label{sdef}
A_1&=&\frac{(1+\frac{c_3}{r^2})^{\frac{1}{2}}}{(H_1
  H_2)^{\frac{1}{4}}}, \,\,\,\ A_2= r (H_1 H_2)^{\frac{1}{4}}, \,\,\,\
A_3=0\,,\nn \\
K_1&=&\frac{c_3 \coth\sigma}{r^2(H_1
  H_2)^{\frac{1}{4}}(1+\frac{c_3}{r^2})^{\frac{1}{2}}}dt, \,\,\,\,\
K_2= r (H_1 H_2)^{\frac{1}{4}} \frac{1}{2}\cos\theta d \phi ,
\een

Now following the similar procedure as described above we can write
the space-like \tst\ transformed geometry, and it looks like
(\ref{NSNSTsT}), with $\chi$ as $x_5$ and metric functions as given in
(\ref{sdef}). ${\cal M}$  takes the following form,
\be
{\cal M}=(1+ \gamma^2 r^2 (1+\frac{c_3}{r^2}))^{-1},
\ee
and the explicit form of the background is,
\ben
d\hat s^2_{\rm{str}}&=& {{\cal M}\over \sqrt{H_1 H_2}}\bigg[ -dt^2 +
dx_5^2 + {c_3 \over r^2} ( \coth\sigma dt + dx_5)^2 + \gamma^2
(c_3(\coth^2\sigma-1)-r^2)dt^2\bigg]\nn \\
&+&\sqrt{\frac{H_1}{H_2}} dM_4^2 +
\sqrt{H_1 H_2}\frac{dr^2}{f} + {\cal M}(A_2 d \psi + A_3 d x^- +
K_2)^2+\frac{A_2}{4}(d \theta^2+\sin^2\theta d\phi^2)\,,\nn \\
e^{2 \hat\Phi}&=&{\cal M}\,, \nn \\
\hat B&=&- \gamma {\cal M}\, A_1 A_2\,  dx_5 \wedge (A_2 d \psi + A_3
d\chi + K_2)\, .
\een
We can again study the extremal near horizon limit of this geometry
and we find that it is not independent of the shift parameter
$\gamma$, quite unlike the null \tst\ transformed geometry.  In the
extremal limit, ${\cal M}\rightarrow {\cal M}_0=(1+ \gamma^2
{c_3})^{-1}$, and hence the overall $\gamma$ dependence does
not drop out.  The two form NS-NS field is also non trivial in this
limit.

\subsubsection{\tst\ of Non-extremal $D1$-$D5$ System: BTZ Black Hole
  and $CFT_1$}

In this section we consider the decoupling limit of non-extremal
$D1$-$D5$ system which is $BTZ\times S^3\times T^4$. The boundary of
this $AdS$ black hole is two dimensional. We start with
the metric (\ref{d1d5}) and consider the following decoupling limit\footnote{See \cite{justin} for
detailed discussion}
\ben
\alpha' \to 0, \quad r\to 0, \quad r_0\to 0\,,\nn\\
\alpha, \gamma, \sigma \to \infty
\een
with
\ben
U= \frac r {\alpha'}, \quad U_0 =\frac {r_0}  {\alpha'} = {\rm fixed},
\een
In this limit the metric (\ref{d1d5}) becomes,
\ben
ds^2 &=& \alpha'\ltb \frac{U^2}{l^2} \lb -dt^2 +dx_5^2\rb +\frac
{U_0^2}{l^2} \lb \cosh\sigma dt + \sinh\sigma dx_5\rb^2 + \frac{dU^2}
{U^2-U_0^2} \right.
\nn\\
&& \hspace{4.7cm} \left. + l^2 d\Omega_3^2 + \sqrt{\frac{Q_1}{v Q_5}}
(dx_6^2+\cdots +dx_9^2) \rtb\,.
\een
After performing the following coordinate redefinition,
\ben
u^2 &=&{(U^2 +U_0^2 \sinh^2\sigma) \over l^2}, \quad \tau = l t,\quad %\nn\\
r_+ = \frac{U_0 \cosh\sigma}{l}, \quad r_-=\frac{U_0 \sinh\sigma}{l}\,,
\een
the $BTZ$ black hole metric becomes,
\ben
ds^2 &=& \alpha'\ltb - \frac{u^2}{l^2} f d\tau^2 +u^2 \lb dx_5+ \frac
{r_+ r_-}{u^2 l}d\tau \rb^2  + \frac{l^2 du^2}
{u^2 f} \right.
\nn\\
&& \hspace{1.7cm} \left. + l^2 d\Omega_3^2 + \sqrt{\frac{Q_1}{v Q_5}}
(dx_6^2+\cdots +dx_9^2) \rtb\,,
\een
where different quantities appearing in the metric are,
\ben
v&=& \frac{V_4}{16 \pi \alpha'^2}={\rm fixed},\quad l=(g_6^2 Q_1
Q_5)^{1/4},\quad g_6 = \frac{g_s}{\sqrt v}\,,\nn\\
f&=& \lb 1-\frac{r_+}{u^2}\rb\lb 1-\frac{r_-}{u^2}\rb\,,
\een
where, $V_4$ is the volume of $T^4$ and $g_s$ is the string coupling.
We will carry out only null \tst\ transformation on the BTZ black hole
solution.  As usual, we will again define light-cone coordinates as
$x^{\pm}=\frac{1}{2}(t \pm l\, x_5)$ and follow the procedure outlined
earlier.  The final form of the \tst\ transformed geometry is,
\ben
d\hat s^2&=&{\cal M}(A_1 dx^-+K_1 dx^+)^2+{\cal M}
\left (A_2 d\psi+\frac{1}{2}\cos\theta d\phi\right )^2-A_4 (dx^{+})^2 \nn \\
&& +\frac{l^2}{4} \left (d\theta+\sin^2\theta d \phi\right )^2 +
\sqrt{\frac{Q_1}{v Q_5}} (dx_6^2+\cdots +dx_9^2), \nn \\
A_1&=&\frac{r_+ -r_-}{l}, \quad A_2 =l, \quad K_1=\frac{r_+^2+ r_-^2-2
  r^2}{l(r_+-r_-)},\nn\\
 A_4 &=&4 \frac{(r^2-r_-^2)(r^2-r_+^2)}{l^2(r_+-r_-)^2}, \quad
{\cal M}=(1+ \gamma^2 (r_+-r_-)^2)^{-1}\, .
\een
This metric has asymptotic Schr\" odinger symmetry.  This can be seen
by carrying out large $r$ asymptotic expansion of the above geometry
and in particular, the asymptotic metric in light-cone coordinates has
the form,
\ben
d\hat s^2 &=& {\cal M} \ltb -\frac{4 \gamma^2 r^4 }{l^2}
(dx^+)^2 - \frac {2r^2}{l^2} dx^+ dx^-\rtb +{\cal M}
(d\psi+\frac{1}{2}\cos\theta d\phi)^2 \nn\\
&& \hspace{4.8cm}+\frac{l^2}{4} \left
  (d\theta+\sin^2\theta d \phi\right )^2  + \frac{l^2 dr^2}{r^2}\,.
\een
It is evident that the asymptotic Schr\" odinger background has the
dynamical exponent $z=2$.

It is worth pointing out at this point that the boundary field theory
in this case is one dimensional.  In other words, this geometry is
dual to quantum mechanical system.  Given the fact that the bulk
geometry asymptotes to Schr\" odinger spacetime with dynamical
exponent $z=2$ implies the boundary theory is conformal quantum
mechanics.  This is because for $z=2$, Schr\" odinger spacetimes have
additional symmetry corresponding to special conformal
transformations.  Recall that conformal quantum mechanics plays
pivotal role in Sen's quantum entropy function\cite{sen}.  It would be
interesting to see if a similar technique could be extended to
non-relativistic black holes.

\section{\tst \ Transformation of Rotating M2 Brane}
\label{M2-brane}

In this section we will look at the rotating M2 brane solutions in the
11-dimensional supergravity (low energy limit of M-theory). These M2
branes can have four angular momenta in the transverse plane and they
source the three form potential in the 11 dimension.  Like in the case
of D3 brane, we will start with a rotating M2 brane solution with all
angular momenta equal.  This background in the decoupling limit takes
the form,
\ben\label{solm2}
ds^2&=&-\frac{f(\rho)}{H^2}dt^2+H^2(f(\rho) d\rho^2 + \rho^2
d\Omega_{2,k}^2)+g^{-2}\sum_{i=1}^4(d
\mu_i^2+\mu_i^2(d\phi_i+\mathcal{A}_i)^2, \nn \\
C^M_3&=& -8g^3 \rho^3 H^3 dt\wedge d\Omega_{2,k}^2+2 \sqrt{q(\mu+ k
  q)}
\sum_i \mu_i^2 d\phi_i \wedge d\Omega_{2,k}^2
\een
where,
\ben
f=k-\frac{\mu}{r}+4 g^2 r^2 H^4, \quad
H= 1+\frac q r\,.
\een
We will now do the following coordinate transformation,
\ben
r=\rho +q,\qquad m=\mu + 2  q k, \qquad Q=\sqrt{q(\mu+ k q)}\,.
\een
{}From now on we will work in $k=0$ limit, which implies, $d\Omega_{2,k}^2=
dy_1^2+dy_2^2$.  We also redefine the $M2$ brane worldvolume coordinates,
$\vec y = 2 g \vec x$, and also redefine the coupling constant as
$g=\frac {\hat g}{2}$
so that we can bring the boundary metric in the form $r^2 (-dt^2 +
d\vec x^2)+ M_7$.  The metric and the three form field strength
expressed in terms of these variables become,
\ben\label{solm2nv}
ds^2&=&-r^2 f(r) dt^2 + \frac{dr^2 }{r^2 f(r)} + \frac {r^2}{\hat g^2}
(dx_1^2+dx_2^2) + 4 \hat g^2 \sum_{i=1}^4 \lb d\mu_i^2 +\mu_i^2
(d\phi_i + \mathcal{A})^2\rb, \nn
\\
C^M_3&=&-\hat g  r^3 dt \wedge dx_1\wedge dx_2 + \frac{2Q}{\hat g^2}
\sum_{i=1}^4 \mu_i^2 d\phi_i \wedge dx_1 \wedge dx_2\,,
\een
and
\be
\mathcal{A}=\frac Q {r\hat g} dt, \,\,\,\,\, f=\hat
g^2-\frac{m}{r^3}+\frac{Q^2}{r^4} \,.
\ee
The $S^7$ metric can be written in the Hopf fibration form,
\be
\sum_{i=1}^4 \lb d\mu_i^2 +\mu_i^2 (d\phi_i + \mathcal{A})^2\rb=
(d\psi + \mathcal{P} +\mathcal{A})^2 + ds_{\mathbb{CP}^3}^2
\ee
where,
\ben
\phi_1 &=& \psi + \frac 1 4 (\xi_1 +\xi_2+\xi_3), \quad \phi_2 = \psi
+ \frac 1 4 (\xi_2 +\xi_3-3\xi_1), \nn \\
\phi_3 &=& \psi + \frac 1 4 (\xi_3 +\xi_1-3\xi_2), \quad \phi_4 = \psi
+ \frac 1 4 (\xi_1 +\xi_2-3\xi_3)\,.
\een
The advantage of this form of the metric is in the convenience of
expressing equal angular momenta in four transverse planes.  They
essentially translate into momentum along $\psi$ direction.  The one
form $\mathcal{P}$ is
\be
\mathcal{P}= \frac 1 4 (d\xi_1+d\xi_2+d\xi_3) -\mu_2^2 d\xi_1 -\mu_3^2
d \xi_2 -\mu_4^2 d\xi_3\,,
\ee
and explicit form of the $\mathbb{CP}^3$ metric is
\ben
ds_{\mathbb{CP}^3}^2 &=& d\alpha^2 + \sin^2\alpha d\beta^2 +
\sin^2\alpha \sin^2\beta d\gamma^2\nn \\
&&+ \cos^2\alpha\sin^2\alpha (d\xi_3 + (d\xi_1 - d\xi_3) \cos^2\beta +
(d\xi_2 - d\xi_3) \cos^2 \gamma \sin^2\beta)^2 \nn\\
&& + \sin^2\alpha \sin^2\beta \cos^2\beta (d\xi_1 - d\xi_3 + (-d\xi_2
+ d\xi_3) \cos^2\gamma)^2 \nn \\
&&+ \cos^2\gamma \sin^2\gamma \sin^2\beta \sin^2\alpha
(d\xi_2-d\xi_3)^2 \,.
\een
In terms of the Hopf fibration, the metric and 3 form fields for the
rotating M2 brane solution are given by,
\ben
ds^2 &=&-r^2 f(r) dt^2 + \frac{dr^2 }{r^2 f(r)} + r^2 (dx_1^2+dx_2^2)
+ 4(d\psi+\mathcal{P}+\mathcal{A})^2 + 4 ds_{\mathbb{CP}^3}^2\,, \nn \\
C^M_3 &=& -r^3 dt\wedge dx_1 \wedge dx_2 + 2 Q (d\psi+\mathcal{P})\wedge
dx_1 \wedge dx_2 \,.
\een
Here for simplicity we have set $\hat g =1$.

Since T-duality is not a symmetry of M-theory, we can carry out the
\tst\ transformation only after re-expressing this solution in terms of
string theory variables.  For this purpose, it is useful to write down
the metric and three form field in terms of vielbeins, which for this
background are,
\ben
e^0 &=& \frac{2 r \sqrt{f}}{\sqrt{1-f}} dx^+, \quad e^1 = {r \over
  \sqrt{1-f}}\ltb (1-f)dx^- -(1+f)dx^+ \rtb\,, \nn \\
e^2 &=& {dr \over r \sqrt{1-f}}, \quad e^3= d\alpha, \quad e^4=
\sin\alpha d\beta,\quad
e^5 = \sin\alpha \sin\beta d\gamma\,,\nn \\
e^6&=& \cos\alpha\sin\alpha (d\xi_3 + (d\xi_1 - d\xi_3) \cos^2\beta +
(d\xi_2 - d\xi_3) \cos^2\
\gamma \sin^2\beta)\,, \nn\\
e^7 &=&  \sin\alpha \sin\beta \cos\beta (d\xi_3 - d\xi_1 + (d\xi_2-
d\xi_3) \cos^2\gamma)\,,\nn \\
e^8 &=& \cos\gamma \sin\gamma \sin\beta \sin\alpha  (d\xi_3-d\xi_2),
\quad e^9 = 2(d\psi+ \mathcal{P} + \mathcal{A}), \quad e^{10}=r dx_2.
\een
where,
\be\label{lcdef}
x^+ = \frac 1 2 (t+x_1), \qquad x^- = \frac 1 2 (t-x_1)
\ee
and,
\be
\frac{l^2}{2} d\mathcal{P}=-\omega_{CP3}=-2 (e^3 \wedge e^6 +e^4
\wedge e^7+e^5 \wedge e^8)\,,
\ee
where, $\omega_{CP3}$ is the K\" ahler form on $\mathbb{CP}^3$.
We can dimensionally reduce the M-theory solution along any isometry
direction and the reduced ten-dimensional geometry is guaranteed to be
a solution of the type IIA theory.  Consider an M-theory solution of the
form,
\ben
ds_M^2&=& ds_{10}^2+ e^{2 \sigma(x^{\mu})}(dx_{10}+A_{\mu}dx^{\mu})^2
,\,\,\,\, x^{\mu}={0,1,,,,9}\nn
\\
C^M_3&=&\frac{1}{3!}C^M_{PQR}dx^P \wedge dx^Q \wedge dx^R\,.
\een
After reducing the solution along $x_{10}$ direction,  we get
a solution to the type IIA equations of motion, where the metric is
expressed in the string frame. The NS-NS sector of the solution looks like,
\ben\label{rules}
ds_{IIA}^2= e^{\sigma(x^{\mu})} ds_{10}^2, \,\,\, \Phi=\frac{3
  \sigma(x^{\mu})}{2}, \,\,\,\ B_{\mu\nu}=C^M_{\mu\nu10},
\een
whereas the R-R sector takes the following form,
\ben
C_{\mu}=A_{\mu}, \,\,\,\,\ C_{\mu\nu\rho}=C^M_{\mu\nu\rho}\,.
\een
We will now apply \tst\ transformation on the type IIA solution and uplift
the transformed geometry again to M-theory.  Thus the transformed
background is a new solution to M-theory.  Notice in case of the
rotating M2 brane background we can perform the
dimensional reduction of M-theory solution, in two different ways.  We
can either reduce along the worldvolume direction, {\em i.e.}, $x_2$
or we can reduce along the transverse direction $\psi$.  These two
procedures give rise to two different solutions of type IIA string
theory.  While in the first reduction we end up with a fundamental
string solution of type IIA, the latter one corresponds to D2 brane
solution and therefore the corresponding type IIA background has
nontrivial R-R sector field.  After getting the TST transformed type IIA
solution, the uplifted M-theory solution, in general, takes the following form,
\ben\label{Mthup}
d\hat s_{M}^2&=&e^{-\frac{2 \hat\Phi}{3}}(d\hat s)^2+e^{\frac{4
    \hat\Phi}{3}}(dx_{10}+\hat C_{1\mu} dx^{\mu})^2 \nn \\
\hat C^M_3&=&\frac{1}{3!}\hat C_{3\mu\nu\rho}dx^{\mu} \wedge dx^{\nu}
\wedge dx^{\rho} + \frac{1}{2} \hat B_{\mu\nu} dx^{\mu} \wedge
dx^{\nu} \wedge dx^{10} .
\een
In the next couple of subsections we will look at these two reductions
and their \tst\ transforms in detail.

{\subsection{Reduction along Worldvolume $x_2$ Direction}
\label{m2x2}

Let us consider that the worldvolume direction $x_2$ is compact.  We
can then dimensionally reduce the M2 brane solution to a ten
dimensional solution of type IIA theory.  This ten dimensional
solution will only have the NS-NS sector fields turned on.  Now
following (\ref{rules}), we can read out the metric, dilaton and NS-NS
two form field,
\ben\label{IIA}
ds_{IIA}^2 &=& r \ltb -r^2 f(r) dt^2 + \frac{dr^2 }{r^2 f(r)} + r^2
dx_1^2 +4 (d\psi+\mathcal{P}+\mathcal{A})^2 +
4 ds_{\mathbb{CP}^3}^2 \rtb\,,\nn \\
B_{\mu\nu} &=& C^M_{\mu\nu x_2} = -r^3 dt\wedge dx_1 +
2Q(d\psi+\mathcal{P})\wedge dx_1\,,\nn\\
\exp[\frac 4 3 \Phi] &=& \exp[2 \sigma] = r^2\,.
\een

The metric and 2 form fields are given by,
\be
ds_{10}^2 = r\lb - (e^0)^2 + (e^1)^2 +(e^2)^2+4((e^3)^2+\cdots
+(e^8)^2) + (e^9)^2\rb\,,
\ee
\ben
B &=& 16 r \lb 1 + {Q^2\over 4r^4} \rb {e^0\wedge e^1 \over
  \sqrt{4 f}} + {2Q \over r \sqrt{1-f}} (\sqrt{f} e^0 +e^1)\,
\wedge e^9 \nn \\
H &=&24 r\sqrt{\frac{1-f}{f}} e^0\wedge e^1 \wedge e^2+\frac{4
  Q}{r \sqrt{1-f}} \omega_{CP3}\wedge (\sqrt{f} e^0 +e^1)\,.
\een

We follow the \tst \ transformation rules given in section 2,
(\ref{eq:4},\ref{eq:1}) and get,
\ben
d\hat s^2 &=& {\cal M} \ltb \lb A_1 dx^- + K_1 dx^+ \rb^2
+ \lb  A_2 d\psi + 2 \sqrt r \mathcal{P} + A_3 dx^- + K_2 dx^+
\rb^2\rtb  \nn\\
&& - A_4 (dx^+)^2 +4 r ds_{\mathbb{CP}^3} +\frac{dr^2}{r f}\,,
\een
\ben
\hat B &=& {\cal M} \bigg[  B (1-2 Q \gamma) \nn\\
&& - \gamma A_1 A_2 \lb A_1 dx^- + K_1 dx^+ \rb
\wedge \lb A_2 d\psi  + 2 \sqrt r \mathcal{P} + A_3 dx^- + K_2 dx^+
\rb \bigg]\,,
\een
where,
\ben
A_1 &=& r\sqrt{r(1-f)} ,
\quad A_2 =  2 \sqrt r, \quad K_1 = - \frac{r^{3/2}(1+f)}{\sqrt{1-f}},
\quad A_3 = K_2= 2 \mathcal{A}_t \sqrt r\,, \nn\\
A_4 &=& \frac{2 \sqrt{r^3 f}}{\sqrt{1 - f}},\quad\hat K_1 = K_1 +
\gamma \frac{4 f Q r^{3/2}}{\sqrt{1-f}},
\quad \hat K_2 = K_2 -\gamma
\frac{4 (2 Q^2 + r^4)}{\sqrt r}\,, \nn \\
{\cal M}&=& (1-4 \gamma Q +4\gamma^2 (Q^2+(1-f)r^4))^{-1}\,.
\een
By carrying out expansion of the metric for large $r$, it is easy to
see that the above solution does not have asymptotic \sch\ geometry. It
is particularly easy to see that ${\cal M}$ does not become constant as we
approach the boundary, it instead vanishes. Thus, although the above
field configuration is a solution of type IIA theory and therefore its
uplift along $x_2$ is a solution in M-theory, it does not possess
either asymptotic AdS or Schr\" odinger symmetry.  We will therefore
not pursue detailed study of this background any further.

We can nevertheless get interesting M-theory solutions with asymptotic
AdS symmetry if we follow a strategy of first performing a chain of
duality transformations and then doing the \tst\ transformation.  In
the process we will have multiple ways of lifting the solution back to
M-theory.  Following will be our strategy of obtaining a new solution
from the type IIA background that we got by reducing rotating M2-brane
solution along $x^2$ direction:
\begin{enumerate}
\item We will first T-dualize the type IIA solution (\ref{IIA}) along
  $\psi$ direction.  This will take us to type IIB theory.  We will
  continue to refer the T-dual direction as $\psi$.
\item Next we perform an S-duality transformation on this solution.
\item We will then carry out the usual \tst\ transformation on this S-dual
solution. The T-duality will be performed on $\psi$ and the shift will
be performed along one of the lightcone directions as
defined in (\ref{lcdef}).
\item We can then go back to type IIA solution, either by
  \begin{enumerate}
  \item performing another S-duality transformation on the \tst\ transformed
  geometry and then T-dualize it back to type IIA or
  \item by directly T-dualizing the deformed geometry to type IIA.
  \end{enumerate}
\item Finally we uplift the resulting type IIA solution obtained by
  either method to M-theory.
\end{enumerate}

We provide details of intermediate steps in appendix
(\ref{m2app}) and present only the final \tst\ transformed solution
uplifted to M-theory here.
If we follow the step $4a$, then the M-theory metric becomes,
\ben
d\hat s_M^2&=&r^2 {\cal M} \big[(1-f-4 f
\gamma^2)(dx^+)^2- 2 (1+f)dx^+dx^- +(1-f)(dx^-)^2 \big] \nn \\
&&+\frac{dr^2}{r^2 f}+4 (d\psi^2 +
 ds_{\mathbb{CP}^3}^2) + e^{\frac{4}{3} \hat\Phi}(dx_2 +
 C^{T2}_{1 \mu} dx^{\mu})^2\,, \nn \\
C_3^M&=&\frac{1}{3!}C^{T2}_{3\mu\nu\rho}dx^{\mu} \wedge dx^{\nu}
\wedge
dx^{\rho} + \frac{1}{2} B^{T2}_{\mu\nu} dx^{\mu} \wedge dx^{\nu}
\wedge dx_{2}\,,
\een
with ${\cal M} =(1+\gamma^2(1-f))^{-1}$.  Expressions for
$C_{1\mu}^{T2}$ and $C_{3\mu\nu\rho}^{T2}$ are given in appendix (\ref{m2app}).

In case we directly T-dualize the \tst\ transformed metric using the
step 4 (b), it gives the following M-theory geometry,
\ben
ds_M^2&=&(1-f)r^2\bigg((dx^-)^2+(dx^+)^2-
2 \frac{1+f}{1-f}dx^- dx^+\bigg)+
\frac{1}{f r^2}dr^2
+{\cal M} r^2d\psi^2 \nn \\
&&+r^2 \gamma ((1+f) dx^+ -(1-f) dx^-)d \psi+
4 ( ds_{\mathbb{CP}^3}^2+(dx_2+C_{1\mu}^{T2} dx^{\mu})^2),
\nn \\
C_3^M&=&\frac{1}{3!}C^{T2}_{3\mu\nu\rho}dx^{\mu} \wedge dx^{\nu}
\wedge
dx^{\rho} + \frac{1}{2} B^{T2}_{\mu\nu} dx^{\mu} \wedge dx^{\nu}
\wedge dx_{2} .
\een

The asymptotic geometry is given by,
\ben
ds^2 &=& \frac{dr^2}{r^2} +r^2 \lb -4dx^+dx^- +4\gamma dx^+ d\psi +
d\psi^2 \rb \nn\\
\hspace{2.5cm} && + 4 \lb dx_2+\sum(-\frac 1 4 + \mu^2_{i+1})d\xi_i\rb^2 + 4
ds_{\cpt}^2\,.
\een

Although the deformed M-theory solution has a non-trivial $\gamma$
dependence and is well defined at the boundary but it is easy to check
that this metric is asymptotically $AdS$.  It does not have a
Schr\" odinger like behavior at the boundary.

\subsection{Reduction along Transverse $\psi$ Direction}

We can also reduce the M-theory solution along $\psi$ direction, the
corresponding type IIA solution has both NS-NS and R-R sector turned
on.  Again following the general rules given in (\ref{rules}), the type
IIA solution that we get is as follows,
\ben\label{iiatr}
ds_{IIA}^2&=& 2(-r^2 f dt^2+\frac{dr^2}{r^2 f}+r^2(dx_1^2+dx_2^2)+4
ds_{\mathbb{CP}^3}^2)\,, \nn \\
B_2&=& 2 Q dx_1 \wedge dx_2, \,\qquad e^{2\Phi}=8\,, \nn \\
C_{\mu}&=& \mathcal{P}+\mathcal{A}, \qquad C_{3}=(-r^3 dt+ 2 Q
\mathcal{P})\wedge dx_1 \wedge dx_2\,.
\een

We will perform \tst\ transformation on this solution. For
this purpose, we will again define light-cone directions as in
(\ref{lcdef}).  In terms of the vielbeins the metric looks as,
\be
ds^2= 2( - (e^0)^2 + (e^1)^2 +(e^2)^2+4((e^3)^2+\cdots +(e^8)^2) +
(e^{10})^2 )\,.
\ee

We also redefine $\xi_1,\xi_2,\xi_3$ in terms of $\chi_1= \xi_3,\,
\chi_2=\xi_1-\xi_2$
and $\chi_3=\xi_1-\xi_3 $.  The $CP_3$ metric can now be written using
the new variables as,
\be
4 ds_{\mathbb{CP}^3}^2= \sin^2 2\alpha (d\chi_1 + N)^2+ ds_5^2,
\ee
where,
\be
N= d\chi_2 \cos^2\beta + d\chi_3 \cos^2 \delta \sin^2\beta\,,
\ee
and $ds_5^2$ is the remaining part of $CP_3$ metric consisting of
$\alpha,\beta,\delta,\chi_2, \chi_3$.

Next we perform T-duality transformation along $\chi_1$ and the shift
along $x^-$ direction. The final \tst\ transformed solution takes the
following form,
\ben
d\hat s^2 &=& 2{\cal M}((e^1)^2+(e^6 +2 \gamma \frac{Q}{r} \sin2\alpha
\, e^{10})^2 )
\nn \\
&+&2(-(e^0)^2
+(e^2)^2+4((e^3)^2+\cdots +(e^5)^2+(e^7)^2 +(e^8)^2)+(e^{10})^2)\,.
\een
In coordinate basis, the full solution looks like,
\ben
d\hat s^2 &=& {\cal M}((A_1 dx^-+ K_1 dx^+)^2+ (A_2
 (d\chi_1 + N) +2 \gamma Q \sin2\alpha \, dx_2)^2-D (dx^+)^2 \nn \\
&+& 2 (\frac{dr^2}{r^2 f}+ r^2 dx_2^2+ ds_5^2),\nn \\
\hat B &=& {\cal M}\left(B -\gamma A_1 A_2(A_1 dx^-+ K_1 dx^+) \wedge
  A_2  (d\chi_1 + N)+ 16 \gamma^2 f Qr^2 \sin^2 2\alpha dx^+ \wedge
  dx_2 \right), \nn \\
e^{2 \hat\Phi} &=&\frac{8}{{\cal M}}, \quad \hat C_1=
\mathcal{P}+\mathcal{A}+2\gamma Q \cos\alpha\, dx_2 \nn \\
\hat C_3 &=& C_3 + (\mathcal{P}+\mathcal{A})\wedge B - \hat C_1\wedge
\hat B\, .
\een
Various functions appearing in the metric and other background fields are,
\ben
A_1 &=& \sqrt{2(1-f)} r, \quad A_2=\sqrt{2} \sin 2\alpha,\nn \\
K_1 &=&-\frac{\sqrt{2}(1+f)r}{\sqrt{1-f}}, \quad A_4
=\frac{2\sqrt{2f}r}{\sqrt{1-f}}\,.
\een
The asymptotic metric (after uplifting to 11 dimensions) is given by,
\ben
ds^2 &=& \lb -16 \gamma^2 \sin^2 2\alpha \ r^4 (dx^+)^2 -4 r^2 dx^+
dx^- + r^2 dx_2^2 +\frac{dr^2}{r^2} \rb \nn\\
&& + \frac 1 2 (A_2
 (d\chi_1 + N) +2 \gamma Q \sin2\alpha \, dx_2)^2 + ds_5^2+4(d\psi
 +\hat C_1)^2\,.
\een
This metric has an interesting feature, that the $g_{++}$ component of
it depends on $\alpha$, which is one of the $\mathbb{CP}^3$
coordinates.  As a result the four dimensional space is non-trivially
fibered over the periodic coordinate $\alpha$.  Interestingly the
$\alpha$ dependence of $g_{++}$ is such that for $\alpha = 0,\ \pi/2$
and $\pi$ the asymptotic metric reduces to pure AdS$_4$ spacetime,
whereas for all other values of $\alpha$ we get asymptotically Schr\" odinger
background.  We will get back to these special values of $\alpha$ in
the concluding section.  Defining rescaled light-cone coordinates as,
\be\label{m2}
u= 2 \nu  x^+ ,\qquad v=\frac{1}{\nu }x^- , \quad \nu=2 \gamma \sin
2\alpha \,,
\ee
it is easy to see that the above metric takes the form (\ref{sch}) at
the boundary $r \to \infty$ with dynamical exponent $z=2$.  The fact
that the asymptotic geometry is nontrivially fibered over a compact
coordinate of $\mathbb{CP}^3$ has interesting implications for
thermodynamics.  We will find it convenient to isolate the coordinate
$\alpha$ from rest of the $\mathbb{CP}^3$ coordinates and club it with
the noncompact coordinates in order to give unambiguous definitions of
thermodynamic quantities.  In particular, we will leave out the
$\alpha$ direction and integrate out rest of the $\mathbb{CP}^3$
coordinates along with the Hopf fiber coordinate $\psi$.  This will
leave us with a five dimensional spacetime, which as mentioned above
looks asymptotically like a four dimensional Schr\" odinger space
fibered over $\alpha$ circle.  All the thermodynamic quantities
associated with the black hole in the interior of this spacetime will
depend on where we are located in $\alpha$ direction.

%%%%%%%%%%%%%%%%%%%%%%%%%%%%%%%%%%%%%%%%%%%%%%%%%%%%%%%%%%%%%%%%%%%%%%%%%
%%%%%%%%%%%%%%%%%%%%%%%%%%%%%%%%%%%%%%%%%%%%%%%%%%%%%%%%%%%%%%%%%%%%%%%%%

\section{Thermodynamics and Phase Structure}
\label{thermo}

We will study thermodynamics and possible phase transition of
these non relativistic systems to other non-relativistic systems.  To
understand the phase structure of a black hole spacetime one needs to
compute its free energy\cite{cvetic-gubser} which is given by
\be
W=\frac{I}{\beta}\,,
\ee
where, $I$ is the (Euclidean) on-shell action and $\beta$ is the inverse
temperature. In asymptotically $AdS$ space the on-shell action has large
volume divergences\footnote{For asymptotically flat black holes the
  divergence comes from the Gibbons-Hawking boundary term.} and we
need to regularize this action by either using boundary counterterms
\cite{bala-kraus} or subtracting contribution of background
spacetime \cite{hawking-page}.  A detailed discussion of background
subtraction method can be found in \cite{dutta-gopakumar}.

In the Hawking-Page formalism (or background subtraction method), one
has to match the geometry of black hole spacetime and background
spacetime at some constant $r=\tilde R$ hyper-surface. This matching
fixes the temperature of background spacetime in terms of that of the
black hole spacetime, and at the end one takes the cutoff $\tilde R
\ra \infty$.  In \cite{yamada, kim-yamada} it has been argued that for
black holes in asymptotically \sch \ space, the subtraction method is
subtle as the null circle becomes degenerate for background spacetime
on constant $\tilde R$ hyper-surface, whereas for black hole spacetime
it is not. Therefore they introduced an "{\em ad hoc}" prescription to
compute renormalized on-shell action.

It is worth emphasizing that the Euclidean method, by itself, is
sufficient to compute all thermodynamic quantities, which satisfy the
first law, starting from the renormalized on-shell action.  The
on-shell action (or more precisely the free energy) captures phase
structure of the black hole spacetime.  Therefore, our goal here is to
compute the free energy so that we can analyze the phase structure of
these solutions.  While the Euclidean method is sufficient, it can
become quite cumbersome for complicated field configurations.  Our new
solutions with Schr\" odinger asymptotics do indeed have several
fields turned on.  While it is still possible to evaluate the
Euclidean action, we will momentarily define an alternate method for
computing the on-shell Euclidean action.  First of all notice that the
Hawking temperature can be determined either by computing surface
gravity on the horizon or by computing periodicity of the Euclidean
time circle.  In either case, it suffices to have knowledge of the
near horizon geometry.  Secondly, recall entropy of black hole
spacetime can be determined in two different ways.  One way is to
evaluate the Euclidean on-shell action in the black hole spacetime and
then using thermodynamic relations to derive expression for entropy.
Another method is to use Wald's formula to directly derive black hole
entropy.  In all known relativistic examples, it is known that both
these methods give rise to same formula for
entropy\cite{dutta-gopakumar}.  The strategy we will employ rests on
the equality of these two ways of deriving the entropy.  We will first
determine the entropy using Wald's method and turn the Euclidean
action method on its head to derive the on-shell action by integrating
the Wald entropy.  This requires solving a first order inhomogeneous
differential equation.  The on-shell action derived in this manner
contains an ambiguity corresponding to the constant of integration.
This ambiguity can be fixed by demanding that the free energy vanishes
for a fixed background spacetime.  We will subsequently write down the
free energy for the black hole spacetime using this on-shell action
and the Hawking temperature derived from either of the methods
mentioned above.  We will illustrate this method for the $AdS$
Reissner-Nordstr\" om black hole in d+1 dimensions, however, we will
eventually apply it to the asymptotically Schr\" odinger backgrounds.
This method becomes particularly efficient in case of the \tst\
transformed geometers where a variety of fields are turned on due to
solution generating technique.  While the Euclidean action can
nevertheless be determined by conventional methods, our method turns
out to be much more efficient in deriving desired results.

Let us start with $AdS$-\rn \ black hole in d+1 dimensions.  The solution
is given by,
\ben
ds^2 &=& -V(r) dt^2 + \frac{dr^2}{V(r)} + r^2 d\Omega_{k,d-1}^2\,, \\ \nn
A(r) &=& \lb-\frac 1 c \frac q {r^{d-2}} + \mu \rb dt\,,
\een
where,
\be
V(r) = k - \frac{m}{r^{d-2}} + \frac{q^2}{r^{2d -4}} +
\frac{r^2}{L^2}, \qquad c= \sqrt{\frac{2(d-2)}{d-1}}.
\ee
Here $k$ determines the horizon topology.  For flat horizon (black
brane), $k=0$, whereas $k=1(-1)$ for spherical (hyperboloid)
horizons.  The asymptotic value of the gauge field $A_t$ is defined to
be the chemical potential,
\be\label{chpot}
\mu =  \frac 1 c \frac q {r^{d-2}}.
\ee

The temperature of the black hole can be computed from its surface
gravity at the horizon defined as \cite{rangamani},
\be\label{surgra}
\kappa^2=-\frac{1}{2}(\nabla^a\zeta^b)(\nabla_a\zeta_b)|_H,
\ee
where $\zeta^a$ is the null generator of the horizon. The Hawking
temperature is then, $T=\frac{1}{2 \pi} \kappa$. We can also find the
temperature as the inverse of the period of the Euclidean time
circle. Thus the classical solution has a finite temperature
\be\label{temprnbh}
T=\frac 1 {\beta} = \ltb \frac{4 \pi L^2 r_+^{2d-3}}{dr_+^{2d-2} +
  k (d-2) L^2 r_+^{2d-4} -(d-2)q^2 L^2}  \rtb^{-1}\,,
\ee
where $r_+$ is the position of the outer horizon.

We compute entropy using Wald's formula.  For two derivative gravity
theory, the formula gives entropy as the horizon area ${\cal A}_{\rm
  Horizon}$.  To employ Wald's formula, we need the metric to be
written in Einstein's frame.  We find the horizon area
\cite{rangamani-lunin} by writing the metric in the Boyer-Lindquist
coordinate, so that there is a r-coordinate which does not mix with
others\footnote{All the metrics that we have considered in this paper
  are in the Boyer-Lindquist coordinates.}.  We can then write the
metric as
\be\label{en1}
g_{\mu\nu}=\left(\begin{array}{c|c c c}g_{rr} & & 0 & \\ \hline \\ 0 &
    & P_{ij}\\ & \end{array}\rb\,.
\ee
Note that the time coordinate $t$ appears in the metric $P_{ij}$.  The local
definition of the horizon is a fixed $(t,r)$ surface where $g_{rr}$
diverges and the determinant of $P_{ij}$ vanishes.  The cofactor of
$P_{ij}$ is,
\be\label{en2}
A^{\mu\nu}=g^{\mu\nu} det P\,.
\ee
The area of the horizon can then be written as ${\cal A}_{\rm Horizon}=
\sqrt{A^{tt}}$.  Using this formulation, we get entropy
\be\label{entropyrnbh}
S= \frac{{\cal A}_{\rm Horizon}}{4 G} = \frac{V_{k,d-1} r_+^{d-1}}{4G},
\ee
where, $V_{k,d-1}$ is the volume of the transverse space.  In the next
subsection we will use thermodynamic relations to derive the on-shell
action using the expression for entropy given in (\ref{entropyrnbh}).

\vspace{4mm}
\subsection{Computation of On-Shell Action}
\label{sec:comp-shell-acti}
\vspace{2mm}

We will now use the expressions derived above for the entropy and the
temperature and substitute them in thermodynamics relations to compute
the on-shell action or equivalently the free energy.  The expression
for free energy depends on the ensemble we are working on.  In the
present situation we have two ensembles: fixed potential ensemble and
fixed charge ensemble\cite{rob, 10auth}.

\vspace{4mm}
\noindent
{\bf Fixed Potential case}\\

In this ensemble the relation between entropy and free energy is given by,
\be\label{Ieqnfixpot}
S-\beta \lb \frac{\partial I}{\partial \beta}\rb_{\mu} + I =0
\ee
Once we determine the on-shell action $I$ we can compute other
thermodynamic quantities, like the energy and the physical charge using
\ben\label{thermorelationfixpot}
E&=&\lb \frac{\partial I}{\partial \beta}\rb_{\mu} -{\mu \over \beta}
\lb \frac{\partial I}{\partial \mu}\rb_{\beta}\,, \nn \\
Q&=&  -{1 \over \beta} \lb \frac{\partial I}{\partial \mu}\rb_{\beta}.
\een

Now let us solve equation (\ref{Ieqnfixpot}) using (\ref{temprnbh})
and (\ref{entropyrnbh}).  The solution is given by,
\be
I = -\frac{V_{k,d-1}\beta }{16 \pi G L^2}\lb L^2 r_+^{d-2}(c^2 \mu^2
-k)+( r_+^n - 4 c_1)\rb .
\ee

However as we pointed earlier, we have an undetermined constant $c_1$,
which has to be fixed.  This is achieved by employing the following
strategy: the free energy of the black hole space-time is measured
with respect to ``some'' background space-time.  We, therefore, choose
the scale such that the free energy of the background is zero.  With
this choice, we can fix the constant $c_1$.  If we want to measure
the free energy of the black hole spacetime with respect to global
$AdS$ then we set the free energy of the global $AdS$ spacetime to be
{\em zero}.  The global $AdS$ spacetime corresponds to $r_+=0$ keeping
$\mu$ fix in this ensemble\footnote {For black-branes, we chose the
  background to be $AdS$ soliton and set the free energy of the
  soliton to zero.}. This implies
\be
W = \frac{I}{\beta}{\bigg |}_{r_+\ra 0} = 0.
\ee
This uniquely fixes $c_1 =0$.  Therefore the on-shell action is given by,
\be
I = -\frac{V_{k,d-1}\beta }{16 \pi G L^2}\lb L^2 r_+^{d-2}(c^2 \mu^2
-k)+ r_+^n\rb .
\ee
This result was obtained in \cite{rob} and
the complete phase structure of the  $AdS$-\rn \ black hole was
analyzed there\footnote{Also see \cite{shibaji} for black brane phase structure.}.

\vspace{4mm}
\noindent
{\bf Fixed Charge case}\\

One can also consider thermodynamic ensemble where the charge
parameter $q$ is fixed.  In this case the thermodynamic relations are
given by,

\be\label{Ieqnfixcharge}
S-\beta \lb \frac{\partial I}{\partial \beta}\rb_{q} + I =0\,,
\ee
and
\ben\label{thermorelationfixcharge}
E&=&\lb \frac{\partial I}{\partial \beta}\rb_{q}, \nn \\
\mu &=&  {1 \over \beta} \lb \frac{\partial I}{\partial \mu}\rb_{\beta}.
\een

We use (\ref{Ieqnfixcharge}) to compute $I$ and it is given by,
\be
I =  \frac{V_{k,d-1}\beta}{16 \pi G L^2} \lb k L^2 r_+^{d-2} - r_+^d +
(2d-3)q^2 L^2 r_+^{2-d} + 4 c_1 \rb .
\ee

For this ensemble,  we can not measure the free energy of the black hole
spacetime with respect to global $AdS$ as the later has zero
charge.  We will instead measure the free energy with respect to the
extremal solution.  The extremal solution has same charge q as the
black hole, whose free energy we wish to derive and its horizon is
located at $r_e$ where,
\be
d\ r_e^{2+2d}+L^2 (d-2)(q^2 r_e^4 - k r_e^2d)=0.
\ee
In this case we are comparing the free energy of the black hole with
that of the extremal solution.  The constant of integration, $c_1$, is
thus determined by setting the free energy for the extremal black hole
to zero.
\be
W=\frac{I}{\beta}{\bigg |}_{r_+\ra r_e, q \ {\rm fix}} =0\,
\ee
gives,
\be
c_1= \frac 1 4 r_e^{-d-2} \lb r_e^{2d +2} -L^2 ((2d-3)q^2 r_e^4 + k
r_e^{2d})\rb.
\ee
The Euclidean action for the black hole is therefore given by,
\be
I = \frac{V_{k,d-1}\beta}{16 \pi G L^2}\lb k L^2 r_+^{d-2} - r_+^d +
(2d-3)q^2 L^2 r_+^{2-d} -\frac{2k(d-1)} d L^2 r_e^{d-2} -
\frac{2(d-1)^2} d \frac{q^2 L^2}{r_e^{d-2}} \rb\,.
\ee
This result was also found in \cite{rob}.  For $k=0$ (black brane)
case, we see that the black brane geometry is always dominant over the
background geometry.  Thus we see that our strategy gives the desired
result in case of $AdS$-\rn\ black hole in d+1 dimensions.  In the
next subsection we will use this strategy to study thermodynamics of
non-relativistic black objects.

\vspace{4mm}
\subsection{Thermodynamics of Non-Relativistic Black Objects}
\vspace{2mm}

We will use our strategy to do similar computations but this time for
$(d+1)$ dimensional non-relativistic black objects (black branes and
black holes) and study their phase structure.  Such black objects can
be obtained by dimensional reduction of various non-relativistic
branes (D3, M2) along the transverse space.  As we have seen in the
last subsection, for studying the phase structure of a black hole
space-time, we need to know its free energy.  Using thermodynamic
relations, we can obtain the free energy from the knowledge of the
temperature and the entropy of the black hole.

To find the temperature of the non-relativistic black objects, we can
use the surface gravity defined in (\ref{surgra}).  The null generator
$\zeta$ is proportional to $\frac{\partial}{\partial t}$.  In analogy
with asymptotically flat space, where the null generator has unit norm
asymptotically, we can fix the normalization of $\zeta$ . We demand
that the component of $\zeta$ along the boundary (non-relativistic
CFT) time-translation has unit coefficient.  Since the generator of
time translation in the boundary theory is $\frac{\partial}{\partial
  u}$ we get
\be\label{zeta}
\zeta=\frac{1}{\nu}\frac{\partial}{\partial
  t}=\frac{\partial}{\partial u}+\frac{1}{2 \nu^2}\frac{\partial}{\partial v}\,,
\ee
where, $\nu$ is the scaling parameter of the light-cone directions,
which is required for obtaining the boundary $r \to \infty$ metric as in
(\ref{sch}).  In particular, $\nu=\gamma l$ for non-relativistic
rotating D3 brane geometry  (\ref{d3}) and $\nu=2 \gamma \sin 2\alpha$
for non-relativistic rotating M2 brane geometry  (\ref{m2}).  We can
also find the temperature as the inverse of the period of the
Euclidean time circle.  Either of these procedures give us,
\be\label{tempnr}
T=\frac 1 {\beta} = \ltb \frac{4 \pi r_+^{2d-3} \nu }{dr_+^{2d-2}
  +k (d-2) r_+^{2d-4} -(d-2)q^2 }  \rtb^{-1}\,.
\ee
To compute the entropy, we need to compute the horizon area ${\cal
  A}_{\rm Horizon}$.  We can use (\ref{en1}) and (\ref{en2}),
provided we keep in mind that the correct time direction for the
non-relativistic black objects is $u$.  This gives us the entropy as,
\be
\label{entropynr}
S=\frac{{\cal A}_{\rm Horizon} }{4 G} = \frac{V_{k,d-1}
  r_+^{d-1} \nu }{4G}.
\ee

\vspace{4mm}
\subsubsection{On-Shell Action for Fixed Potential Case}
\vspace{2mm}

To study the phase structure, we have to first decide on the choice of
ensemble.  Now that the non-relativistic theories can be realized as a
deformed version of relativistic quantum field theories, it is natural
to allow the particle number to fluctuate.  Thus grand canonical
ensemble is the right choice of ensemble in this case.  Hence we study
the phase structure of non-relativistic systems only for the fixed
potential ensemble.  The strategy described above can be readily
applied to the fixed charge case, however, we will not pursue that
line here.  We hope to report on that elsewhere.

To compute the on-shell action, we use the temperature and entropy as
input and use relation (\ref{Ieqnfixpot}).  In this case, we have two
chemical potentials corresponding to two conserved charges.
As the killing generator $\zeta$ of the event horizon has a component
along the light-like direction $v$ (\ref{zeta}), the corresponding
momentum $\frac{\partial}{\partial v}$ is a conserved charge and we
have a chemical potential for $v$-translation given as,
\be
\mu_1= \frac 1 {2 \nu^2}.
\ee
The other chemical potential is associated with the boundary value of
the gauge field (\ref{chpot}). With $u$ being the correct boundary time
direction, we decompose $A=A_t dt= A_u du+ A_v dv$.  Thus the second
chemical potential is defined as,
\be
\mu_2= \lim_{r \ra \infty}A_u=\frac 1 {2 \nu c} \frac q {r_+^{d-2}} =
\frac{\sqrt{\mu_1}}{\sqrt{2}c}  \frac q {r_+^{d-2}}\,.
\ee
Replacing the charges in-terms of the corresponding chemical
potentials, the temperature in grand canonical ensemble is given by,
\be\label{Temp}
\frac 1 T = \beta =\frac{2 \sqrt{2} \pi  r_+ \sqrt{\mu _1}}{-2 c^2 d
  \mu _2^2+4 c^2 \mu_2^2+(d-2) k \mu _1+d r_+^2 \mu _1}.
\ee
We can then solve the eq.(\ref{Ieqnfixpot}) to write the on-shell action as,
\be\label{ac}
I=\frac{-2 \sqrt{2} c^2 \mu _2^2 r_+^d+8 c_1 r_+^2 \sqrt{\mu _1}+\sqrt{2} k
   \mu _1 r_+^d-\sqrt{2} \mu _1 r_+^{d+2}}{8 r_+ \sqrt{\mu _1} \left(-2
   c^2 d \mu _2^2+4 c^2 \mu _2^2+(d-2) k \mu _1+d r_+^2 \mu _1\right)}\,.
\ee
The corresponding free energy is given by,
\be\label{FE}
W=\frac{r_+^d \left(k \mu _1-2 c^2 \mu _2^2\right)+4 \sqrt{2} c_1 r_+^2
   \sqrt{\mu _1}-\mu _1 r_+^{d+2}}{16 \pi  r_+^2 \mu _1}\,.
\ee
As in the relativistic case, we have an undetermined constant $c_1$
that needs to be fixed.  Choosing the non-relativistic extremal black hole/brane
geometry\footnote{This geometry is obtained by setting $r_+ \ra 0$ and
$Q\ra 0$ of the higher dimensional brane solution.} as the background and
setting its free energy to zero, we get $c_1\ra 0$  and the free energy
becomes,
\be\label{FE2}
W=\frac{r_+^d \left(k \mu _1-2 c^2 \mu _2^2\right)-\mu _1
  r_+^{d+2}}{16 \pi  r_+^2 \mu _1}.
\ee
Thus, from (\ref{FE2}) we see that black branes (k=0) are always
dominant\footnote{The black brane can have a transition to the AdS soliton
geometry\cite{imeroni-sinha, NS}} whereas, there exists a phase
transition between the black hole (k=1) phase and the background
spacetime when
\be\label{cond}
\frac{\mu_1}{\mu_2^2} = \frac{2 c^2}{1-r_+^2}.
\ee
When the ratio ${\mu_1 \over \mu_2^2}$ is less than its critical value (\ref{cond})
the black hole geometry will be dominant over the
background. This is the usual Hawking-Page transition.
Also, if we assume that both  the chemical potentials are positive, then
$r_+ <1$.

It is clear from the equation of $\beta$ (\ref{Temp}), that there are two
distinct behaviour of $\beta$ as a function of the ratio of two
chemical potentials, namely $\mu_1/\mu_2^2$.  For
$\frac{\mu_1}{\mu_2^2}< 2 c^2,\ \beta$
diverges when\footnote{we have considered $d>2$.} $r_+=\frac{(d-2)(2
  c^2 \mu_2^2-\mu_1)}{d \mu_1}$, whereas it smoothly goes to zero as $r_+
\ra 0$ for $\frac{\mu_1}{\mu_2^2} > 2 c^2$.  In the first case there exists
a unique black hole associated with each temperature and this branch
dominates the thermodynamics (free energy is always negative).  In the
second case, for a fixed chemical potentials there exists a nucleation
temperature
\be
\frac 1 {T_n} = \beta_n =
\sqrt{\frac{2 \pi^2}{d(d-2)(\mu_1-2c^2 \mu_2^2)}},
\ee
at which two black holes with same horizon radii are formed with
$r_{n}=\sqrt{\frac{(d-2)(\mu_1-2c^2 \mu_2^2)}{d \mu_1}}$.  As we
increase the temperature one of them becomes smaller (small black
hole) and the other one becomes larger (big black hole).  For
temperatures greater that $T_N$, these two black holes have horizon radius
$r_+ = \lambda r_{n}$ where, $\lambda>1$ for big black hole and
$\lambda<1$ for small black hole.  We compute free energies for these
two black holes and it turns out that free energy for small black hole
is always positive and therefore this phase is unstable whereas the
big black hole phase is dominant over background spacetime for
$\lambda>\sqrt{\frac d {d-2}}$ which is compatible with
(\ref{cond}).  The Hawking-Page transition temperature is,
\be
T_{HP} = (d-1)\sqrt{\frac{\mu_1-2c^2 \mu_2^2}{2 \pi^2}}.
\ee
We can also
find conserved charges corresponding to the chemical
potentials.  They are given by\footnote{We have rewrite $r_+$ in-terms
  of temperature as
\be
r_+ = \frac{\sqrt{2} \pi T \sqrt{\mu
      _1}-\sqrt{\mu _1 \left(2 \pi ^2 T^2-(d-2) d \left(k \mu _1-2 c^2
          \mu _2^2\right)\right)}}{d \mu _1},
\ee
we choose this
  branch as $T \to \infty$ when $r_+ \to 0$.},
\ben
P_1&=& - \frac {\partial W} {\partial \mu_1} \bigg|_{T,\mu_2} =
-\frac{\nu ^2 r_+^{-d-2} \left(d \left(r_+^{2 d}
   \left(k+r_+^2\right)+q^2 r_+^4\right)-2 k r_+^{2 d}\right)}{16 \pi
}\,, \nn \\
P_2&=& - \frac {\partial W} {\partial \mu_2} \bigg|_{T,\mu_1} =
\frac{\sqrt{2(d-2) (d-1)} q \nu }{4\pi }\,.
\een
These are the physical charges for generic $(d+1)$ dimensional
non-relativistic black objects. In particular for 5 dimensional charged
non-relativistic black branes, above results match exactly with
those of \cite{imeroni-sinha}\footnote{Please note that the chemical
  potential $\mu_2$ is $\frac{1}{c}$ times their $\mu_2$.}.

The charge $P_2$ obtained for the non-relativistic case is $2 \nu$
times that of the relativistic charge of \cite{rob}. This is easily
understood from the fact that the gauge field here is $A_u$.

\section{Discussion}
\label{sec:discussion}

We analyzed \tst\ transformed geometries associated with brane
configurations in type II string theory as well as in M-theory.  One
of these configurations was studied earlier in the literature.  We
find that the space-like \tst\ transformation in the rotating D3 brane
case commutes with the extremal near horizon limit.  The geometry
obtained by \tst\ transforming the $D1$-$D5$-$p$ solution has the
feature that in the extremal near horizon limit it reduces to the
undeformed extremal near horizon geometry.  The reason is that in this
limit $\gamma$ dependence drops out completely, which essentially
erases the memory of the shift transformation.  In case of the
BTZ black holes, the dual field theory is non-relativistic conformal
quantum mechanics.

The \tst\ transformation of M2 brane geometry is carried out by using
a set of U-duality transformations.  We carry out these
transformations by first reducing the system to type II set up and
then doing \tst\ and lift it back to the M-theory solution.  Among
various ways of generating new M-theory solution, we find that the
null \tst\ transformation on the type IIA background obtained by
reduction along $\psi$ direction has the feature that the metric is
singular if one of the compact coordinate $\alpha = 0,\ \pi/2,\ \pi$.
Asymptotically this metric approaches Schr\" odinger metric for
generic $\alpha$ but for these special values of $\alpha = 0,\ \pi/2,\
\pi$, the space becomes AdS.  However, for precisely these values of
$\alpha$ the metric has curvature singularity.  One way to get around
this is to consider blow ups of $\mathbb{CP}^3$ at these points so
that the singularity is removed.  These spaces are $\mathbb{CP}^3$
analogs of del Pezzo spaces and the M-theory solution is well behaved
on such blown up spaces.

We also analyzed thermodynamics of these geometries.  We have taken a
novel approach to derive free energy of the system and our results
match with known results in the literature both for fixed charge and
fixed chemical potential ensemble.  Using the expression for free
energy thus derived we have analyzed phase structure of generic
non-relativistic $d+1$-dimensional black objects (holes and branes).
Since these non-relativistic systems are derived from the relativistic
ones, it is natural to choose the ensemble which allows changing
particle number, thus the grand canonical ensemble is a natural choice
in this case.  We have shown that for a specific range of chemical
potentials (\ref{cond}), there exist a Hawking-Page transition from
the non-relativistic black hole to non-relativistic extremal brane
geometries.  This thermodynamic analysis can be applied to $d+1$
dimensional systems for $d>2$ only.  In particular, it cannot be used
for the BTZ black hole case.  The entropy and temperature computed for
the BTZ background are real quantities, however, if we compute them
from the action both of them turn out to be imaginary.  While it can
be shown that the temperature is analytic, the entropy is not.  As a
result, knowledge of real temperature and entropy is not sufficient to
deduce the action and hence the free energy uniquely by inverting the
thermodynamic relation.  This is related to the fact that the
Euclidean time circle has imaginary periodicity.

It would be interesting to understand relation of transformed BTZ
solution to non-relativistic conformal quantum mechanics better.  In
particular, extension of quantum entropy function to these cases would
be quite illuminating.  The 1+1 dimensional non-relativistic conformal
field theory obtained from M2 branes may be relevant to the physics of
the Burgers equation which is non-relativistic Navier-Stokes equation in
one spatial dimension.  We hope to address these issues in future.

\vspace{1cm}

{\bf Acknowledgments}: We would like to thank J. David, R. Gopakumar,
P. Kumar, S. Paul Chowdhury, M. Rangamani, S. Raju, A. Sen and
S. Vandoren for useful discussions.  D.P.J. thanks ITF, Utrecht
University for warm hospitality during the course of this work.
N.B. and S.D. would like to thank ISM2011 and HRI for hospitality
during the completion of the work.  Work of N.B. is supported by NWO
Veni grant, the Netherlands.

\vspace{1cm}

\appendix

\noindent{\large\bf Appendix}

\section{T-duality Transformation Rules for NS-NS and R-R Fields}

For self-consistency of this article, in this appendix, we will jot down
the T-duality transformation rules for both NS-NS and R-R
fields.  For details, the reader can referred to original articles
\cite{Buscher,maharana} and \cite{hassan}.  We will however closely follow the
notation of \cite{imeroni}.

{\bf NS-NS fields:}

Let us denote, the T-dualized direction as $\psi$ and other directions
are denoted by $a,b,..$. Transformation of the NS-NS fields, {\em i.e.},
the metric, the NS-NS two forms $B$ and the dilaton under T-duality is
as follows,
\ben
g'_{\psi\psi}&=&\frac{1}{g_{\psi\psi}}, \qquad
g'_{a\psi}=\frac{B_{a\psi}}{g_{\psi\psi}}, \qquad
g'_{ab}=g_{ab}-\frac{g_{a\psi}g_{\psi b}+B_{a\psi}B_{\psi
    b}}{g_{\psi\psi}} \nn \\
B'_{a\psi}&=&\frac{g_{a\psi}}{g_{\psi \psi}}, \quad
B'_{ab}=B_{ab}-\frac{g_{a\psi}B_{\psi b}+B_{a\psi}g_{\psi
    b}}{g_{\psi\psi}}, \quad \Phi'=\Phi-\frac{1}{2}\ln g_{\psi\psi}.
\een

{\bf R-R fields:}

For stating the transformations of different R-R field, we first
define some notation.  First, we decompose a $p$-form ${\cal N}_p$ as,
\be
{\cal N}_p=\bar {\cal N}_p+ {\cal N}_{p[\psi]} \wedge d \psi,
\ee
where, $\bar {\cal N}_p$ does not contain any $\psi$ component and
${\cal N}_{p[\psi]}$ is a $(p-1)$ formed defined as,
\be
({\cal  N}_{p[\psi]})_{a_1....a_{p-1}} =( {\cal
  N}_{p})_{a_1.....a_{p-1}\psi}.
\ee
Next step is to define one forms,
\be
J=\frac{g_{a\psi}}{g_{\psi\psi}}dx^a, \qquad b=B_{{[\psi]}} +
d\psi\,.
\ee
With these definitions, we are ready to state the T-duality
transformation rule for the R-R fields.  They are,
\be
C'_p=C_{p+1[\psi]}+\bar C_{p-1}\wedge b + C_{p-1[\psi]} \wedge b
\wedge J,
\ee
and similar transformations for the R-R field strengths, ${\cal  F}_p$.

\section{TsT Transformed Metrics}\label{m2app}

We have laid down the strategy that we follow in the applying \tst\
transformation to M-theory solution in subsection (\ref{m2x2}).  Here we
will give intermediate steps of those set of duality transformations on
rotating M2-brane solution.

The solution after step 1 is,
\ben
ds_{T1}^2&=& r \ltb -r^2 f(r) dt^2 + \frac{dr^2 }{r^2 f(r)} +4
ds_{\mathbb{CP}^3}^2 \rtb + r^3(1+\frac{Q^2}{r^4}) dx_1^2
+\frac{1}{4r}d\psi^2\,,\nn \\
B^{T1}&=&-(r^3+2 \mathcal{A}_t Q)dt \wedge dx_1+{\cal A}_t dt \wedge d\psi
+\sum_{i=1}^{3}(\frac{1}{4}-\mu_{i+1}^2)d\xi_i \wedge d\psi\,, \nn \\
\Phi^{T1}&=&\Phi-\frac{1}{2}\ln4r=\ln\frac{r}{2}\,, \quad
C^{T1}_0=C^{T1}_2=C^{T1}_4=0\,. \nn
\\
\een
This is a solution in type IIB theory.  The string coupling is still
divergent in the large $r$ limit.  We will use the S-duality symmetry
of type IIB theory to transform this solution to a dual frame.  The
S-duality transformation leaves the metric in Einstein frame and the
self-dual four form field invariant and transforms the dilaton-axion
as well as the NS-NS and R-R two form fields.  Note that the string
frame metric does transform under S-duality,
\ben
ds_{S1}^2&=& e^{-\Phi^{T1}} ds_{T1}^2 \quad
e^{\Phi^{S1}}=e^{-\Phi^{T1}}=\frac{2}{r} \nn\\
B^{S1}&=&C^{T1}_2=0 \quad C^{S1}_2=-B^{T1} \quad C^{S1}_{4}=C^{T1}_4=0
\een
In the S-dual frame the string coupling becomes weak in the large $r$
limit.  We then apply the \tst\  transformation on the above solution
after writing it in terms of the light cone coordinates defined in
(\ref{lcdef}).  Using (\ref{NSNSTsT}), we get the result,
\ben
d\hat s^2&=&{\cal M}(A_1 dx^-+ K_1 dx^+)^2+{\cal M}(A_2 d\psi +A_5
dx^- +A_6 dx^+)^2-A_4 (dx^{+})^2+2 (4
ds_{\mathbb{CP}^3}^2+\frac{dr^2}{r^2 f} ) \nn \\
\hat B &=& - \gamma A_1 A_2 {\cal M} (A_1 dx^-+ K_1 dx^+) \wedge (A_2
d\psi +A_5 dx^- + A_6 dx^+)
\nn \\
e^{2\hat \Phi}&=& e^{2\Phi^{S1}}{\cal M} \quad \hat C_2=C^{S1}_2,
\quad \hat C_0 = \gamma {\cal A}_t,
\quad {\cal M}=(1+\gamma^2(1-f))^{-1},
\een
where remaining R-R sector fields vanish, and
\ben
A_1&=&r \sqrt{2(1-f)} \quad K_1=-\frac{\sqrt{2} r(1+f)}{\sqrt{1-f}} \nn
\\
A_2&=&\frac{1}{\sqrt{2}r} \quad A_5=-A_6=\frac{\sqrt{2}Q}{r} \quad
A_4=\frac{8 f r^2}{1-f}
\een
We then follow step 4 and 5 to get the \tst\ transformed solution in
type IIA theory and finally to M-theory.  As mentioned in our
strategy, there are two way to arrive at the solution in the type IIA
frame.  The solution after step 4(a) is,
\ben
ds_{T2}^2&=&e^{\frac{2}{3}\hat\Phi} \big [ r^2 {\cal M}((1-f-4 f
\gamma^2)(dx^+)^2- 2 (1+f)dx^+dx^- +(1-f)(dx^-)^2) \nn \\
&&+\frac{dr^2}{r^2 f}+4 (d\psi^2 +
 ds_{\mathbb{CP}^3}^2) \big ] \nn \\
B^{T2}&=& -2 Q (dx^++dx^-) \wedge d\psi, \nn\\
C^{T2}_1&=&C^{S2}_{2[\psi]}+C_0\wedge \tilde b =\gamma
{\cal M} A_1 A^2_2  (A_1 dx^-+ K1 dx^+) +\gamma {\cal A}_t \tilde b\nn \\
C^{T2}_{3}&=& -2(r^3 + 2 {\cal A}_t Q) dx^+\wedge dx^-\wedge \bar b+
\hat C_{2[\psi]} \wedge \tilde b
\wedge \tilde J ,
\qquad C^{T2}_5=0
\een
where,
\ben
\tilde b= -\mathcal{A}_t (dx^++dx^-)
-\sum_{i=1}^{3}(\frac{1}{4}-\mu_{i+1}^2)d\xi_i+d\psi,\quad \tilde J =
\frac{A_5 dx^- + A_6 dx^+}{ A_2 }.
\een
On the other hand, if we directly T-dualize the \tst\ transformed metric, as stated
in step 4(b), it gives the following geometry,
\ben
ds_{T2}^2&=&2(1-f)r^2\bigg((dx^-)^2+(dx^+)^2- 2 \frac{1+f}{1-f}dx^-
dx^+\bigg)
+2{\cal M}^{-1} r^2d\psi^2 \nn \\
&&+2r^2 \gamma ((1+f) dx^+ -(1-f) dx^-)d \psi+
\frac{2}{f r^2}dr^2+ 8  ds_{\mathbb{CP}^3}^2, \nn \\
\Phi_{T2}&=& \frac{3}{2} \ln 2, \quad B^{T2}= 2 Q(dx^- - dx^+) \wedge
d
\psi, \nn \\
C_1^{T2}&=&\hat C_{2[\psi]}+\hat C_0\wedge b=-\mathcal{A}_t
(dx^++dx^-)
-\sum_{i=1}^{3}(\frac{1}{4}-\mu_{i+1}^2)d\xi_i +\gamma {\cal A}_t b, \nn \\
C_3^{T2}&=&-2(r^3 + 2 {\cal A}_t Q) dx^+\wedge dx^-\ \wedge b+ \hat
C_{2[\psi]} \wedge b \wedge J,
\een
where,
\be
 b= - \gamma A_1 A_2^2 {\cal M} (A_1 dx^- + K_1 dx^+)+
d \psi, \quad  J= \frac{A_5 dx^- + A_6 dx^+}{ A_2 }. \ee
Finally, we uplift these solutions to M-theory.  The corresponding
M-theory geometry is given in section (\ref{M2-brane}).

%% \newpage

\end{document}